\documentclass[letterpaper,usenatbib,useAMS,usegraphic]{mnras}
\usepackage{graphicx}
\usepackage{multirow}
\usepackage{color}
\usepackage{lscape}
\usepackage[figuresright]{rotating}
\usepackage{longtable}
\usepackage{mathrsfs,amsmath}
\usepackage{epstopdf}
\newsavebox{\tablebox}
\usepackage{hyperref}
\usepackage{times}
\usepackage{longtable,lscape}
\usepackage{natbib}
\usepackage{mathrsfs}
\usepackage{times,xspace,amssymb}
\usepackage{url}
\usepackage{gensymb}
\usepackage{hyperref}
\hypersetup{colorlinks=True, linkcolor=deepluminous, citecolor=divineblue}
\definecolor{divineblue}{rgb}{0.11,0.22,0.73}
\definecolor{deepluminous}{rgb}{0.0,0.47,0.75}
\usepackage{enumerate}
\usepackage{CJKutf8}
\usepackage{amsmath,amssymb}

\newcommand{\lv}{\ifmmode L_{5100} \else $L_{5100}$\ \fi}
\newcommand{\kms}{\ifmmode {\rm km\ s}^{-1} \else km s$^{-1}$\ \fi}
\newcommand{\ergs}{\ifmmode {\rm erg\ s}^{-1} \else erg s$^{-1}$\ \fi}
\newcommand{\ergc}{\ifmmode {\rm erg\ s^{-1}cm^{-2}} \else $\rm erg\ s^{-1} cm^{-2}$\ \fi}
\newcommand{\lb}{\ifmmode L_{\rm Bol} \else $L_{\rm Bol}$\ \fi}
\newcommand{\ledd}{\ifmmode L_{\rm Edd} \else $L_{\rm Edd}$\ \fi}
\newcommand{\hb}{\ifmmode H\beta \else H$\beta$\ \fi}
\newcommand{\ha}{\ifmmode H\alpha \else H$\alpha$\ \fi}
\newcommand{\oiii}{\ifmmode \rm [O\ \sc{III}] \else $\rm [O\ \sc{III}]$\ \fi}
\newcommand{\nii}{\ifmmode \rm [N\ \sc{II}] \else $\rm [N\ \sc{II}]$\ \fi}
\newcommand{\mgii}{\ifmmode \rm Mg\ \sc{II} \else $\rm Mg\ \sc{II}$\ \fi}
\newcommand{\mbh}{\ifmmode M_{\rm BH}  \else $M_{\rm BH}$\ \fi}
\newcommand{\msun}{M_{\odot}}
\newcommand{\rfe}{\ifmmode R_{\rm Fe} \else $R_{\rm Fe}$\ \fi}
\newcommand{\sst}{\ifmmode \sigma_{\rm \ast}\else $\sigma_{\rm \ast}$\ \fi}
\newcommand{\dhb}{\ifmmode D_{\rm H\beta} \else $D_{\rm H\beta}$\ \fi}
\newcommand{\leddR}{\ifmmode L_{\rm Bol}/L_{\rm Edd} \else $L_{\rm Bol}/L_{\rm Edd}$\ \fi}
\newcommand{\feii}{\ifmmode \rm Fe\ \sc{II} \else $\rm Fe\ \sc{II}$\ \fi}
\newcommand{\mdot}{\ifmmode \dot{\mathscr{M}}  \else $\dot{\mathscr{M}}$\ \fi}
\newcommand{\rhb}{\ifmmode R_{\rm BLR}({\rm H\beta})  \else $R_{\rm BLR}({\rm H\beta})$ \ \fi}
\newcommand{\shb}{\ifmmode \sigma_{\rm H\beta} \else $\sigma_{\rm \hb}$\ \fi}
\newcommand{\RL}{\ifmmode R_{\rm BLR}({\rm H\beta}) - L_{\rm 5100} \else $R_{\rm BLR}({\rm H\beta}) - L_{\rm 5100}$ \ \fi}
\newcommand{\ms}{\ifmmode M_{\rm BH}-\sigma_{\ast} \else $M_{\rm BH}-\sigma_{\ast}$\ \fi}
\newcommand{\bd}{\ifmmode \rm \ha/\hb \else $\rm \ha/\hb$ \fi}

\begin{document}
	
\title[The broad-line Balmer decrement for SDSS RM quasars ]{The variability of  the broad-line  Balmer decrement for quasars from the Sloan Digital Sky Survey Reverberation Mapping}
\author[Y. Ma et al.]{Yan-Song Ma$^{1}$, Shao-Jun Li$^{1}$, Chen-Sheng Gu$^{1}$, Jian-Xia Jiang$^{1}$, Kai-Li Hou$^{1}$, Shu-Hao Qin$^{1}$, Wei-Hao Bian$^{1}$ \thanks{E-mail: whbian@njnu.edu.cn}  \\$^1$School of Physics and Technology, Nanjing Normal University, Nanjing 210023, People's Republic of China\\}
\maketitle
	
\begin{abstract}
Based on the spectral decomposition through a code of PrepSpec, the light curves (spanning 6.5 years in the observed frame) of the broad-line Balmer decrement, i.e., the flux ratio of the broad \ha to the broad  \hb line, are calculated for a sample of 44 Sloan Digital Sky Survey reverberation-mapped quasars ($z<0.53$). 
It is found that the logarithm of the mean broad-line Balmer decrement is 0.62 with a standard deviation of 0.15 dex. The relations between the mean Balmer decrement and the SMBH accretion properties (the luminosity, black hole mass, Eddington ratio, accretion rate) are investigated and no obvious correlations are found. 
It is found that there are 27 quasars ($61\%$) showing strong negative correlations between the Balmer decrement  variance and the continuum variance, i.e., the Balmer decrement would be smaller with larger continuum flux.  Assuming that the dust obscuration leads to the variance in the Balmer decrement and the continuum, an expected slope is $-1/3$, which is not consistent with most of measured slopes.  Using the interpolated cross-correlation function,  the time delays between the  inverse Balmer decrement and the continuum are measured  for 14 quasars with the maximum correlation coefficient larger the 0.6. It suggests that the size corresponding to the Balmer decrement lag extends from the BLR size to the torus size.
\end{abstract}
	
	
\section{INTRODUCTION}
Active galactic nuclei (AGN) are accreting supermassive black holes (SMBHs) at the centres of galaxies. The standard paradigm for AGN assumes that the central SMBH is surrounded by an accretion disc (AD) and broad-line region (BLR), which are surrounded by a dusty torus, and a farther narrow-line region (NLR). 
The NLR/BLR  Balmer decrement (\bd)  is usually used as a tool to investigate the properties of gas and dust in AGN \citep[e.g.,][]{Brinchmann2004, OF2006, Lu2019b}. 

Predicted by Case B recombination (at a typical low electron density of $\rm 10^2~ cm^{-3}$ and temperature of $\rm 10^4 K$ ), an intrinsic Balmer decrement value of 2.86 is suggested  for H II region photoionized by a hot star. For Balmer decrement of AGN, a larger value for the intrinsic decrement is predicted from CLOUDY code, where is  due to the presence of NLR/BLR gas of higher densities , much harder ionizing continuum, lower ionization parameter \citep{OF2006, Netzer2013, HeardGaskell2016}. 

The distributions of Balmer decrement of NLR/BLR were reported for different selected AGN samples \citep[e.g.,][]{Dong2005, Dong2008, HeardGaskell2016, Baron2016, Lu2019b}. For a sample of blue AGN defined by the spectral slopes in order to minimize the effect of dust extinction, the distributions of the intrinsic BLR Balmer decrement was investigated \citep{Dong2008}. Based on this sample of extremely blue AGN, \cite{Gaskell2017} suggested that the intrinsic Balmer is $2.72\pm 0.041$, which is consistent with the Case B value of 2.74 \citep{OF2006}. The small scatter of 0.04 dex suggested the BLR decrement can be used a reddening indicator for the BLR, as well as for the NLR decrement \citep[e.g.,][]{Dong2008, Bian2010a, HeardGaskell2016, Lu2019b}.  For a compiled sample of AGN with reliable decomposition of the narrow/broad \bd lines \citep{Osterbrock1977, Cohen1983, Dong2005, Dong2008}, it was also found that the average BLR Balmer decrement is larger than NLR \citep{HeardGaskell2016}, suggesting an additional dust reddening the BLR is interior to the NLR, which is possibly due to the dust in BLR and/or in torus. However, this is opposite to the results obtained by others,  \citep{d85, Dong2008, Baron2016, Lu2019b}.  The comparison between the NLR and BLR  Balmer decrement depends on the method to de-composite the BLR/ NLR contribution in \ha and \hb lines.

For quasars from the Sloan Digital Sky Survey (SDSS), it was found that BLR Balmer decrement depends on the optical continuum slope, although with a larger scatter \citep[see Fig. 7 in][]{Baron2016}. It was also found that there is no correlation between the decrement and the luminosity \citep{Dong2008, Lu2019b}. Recently the variability of the BLR Balmer decrement is investigated for individual AGN \citep[e.g.,][]{Siram2021, RS2022, Li2022, Selwood2023}. 
The Balmer decrement variability for a sample with spectral monitoring would provide clues to these problems.

It was suggested that the diversity of AGN can be unified by the orientation and the Eddington ratio \citep{Sh14}. It is necessary to investigate the relation between the BLR/NLR decrement and the accretion process.  
For broad-line AGN, the BLR clouds (size $R_{\rm BLR}$ and velocity $\Delta V$) can be used as a probe to calculate its virial mass as follows \citep[e.g.,][]{Pe04, Yu2020b}:
 \begin{equation}
 \label{eq1}
\mbh=f\times \frac{R_{\rm BLR}~(\Delta V)^2}{G},
\end{equation}
where G is the gravitational constant, $f$ is a virial factor. 
From the standard disk model of \cite{SS73}, the dimensionless accretion rate \mdot can be calculated from \lv and \mbh as follows  \citep[e.g.,][]{Du16}: 
 \begin{equation}
 \label{eq2}
\mdot \equiv \dot{M}/\dot{M}_{Edd} = 20.1\left(\frac{\it{l}\mathrm{_{44}}}{\cos\,\it{i}}\right)^{3/2} m_{7}^{-2},
\end{equation}
where $\dot{M}$ is the accretion rate, the Eddington accretion rate $\dot{M}_{Edd}=\ledd/ c^2$, $\l_{44}=\lv/10^{44}\ergs$ is the 5100 \AA\ luminosity in units of $10^{44}$ \ergs, \ledd is the Eddington luminosity, $m_{7}=M_{\rm BH}/10^7 \msun$ is the SMBH mass in units of $10^7 \msun$, $i$ is the accretion disk inclination relative to the observer and $cos i=0.75$. The Eddington ratio \lb/\ledd and the dimensionless accretion rate \mdot are two important parameters describing the SMBH accretion process, where \lb is the bolometric luminosity. Considering $\lb=\eta \dot{M}c^2$, where $\eta$ is the accretion efficiency, $\leddR=\frac{\eta \dot{M}c^2}{\dot{M}_{\rm Edd}c^2}=\eta \mdot$.

In this paper,  for a sample of 44 SDSS-RM quasars, we use light curves (spanning 6.5 years) of \hb, \ha and the continuum at 5100 \AA\ derived from the spectral decomposition through PrepSpec to investigate the variability of BLR Balmer decrement (\bd). In Section 2, we describe the sample and light curve data. The mean value of the Balmer decrement and it connection with the SMBH accretion, the relative variance of the Balmer decrement, the relation of \bd variability with the continuum, and the lag of the Balmer decrement with respect to the continuum are presented in Section 3.  Section 4 is our results.

\section{Sample}

SDSS-RM Project monitored a flux-limited ($i_{\rm psf} < 21.7$) sample of 849 quasars with redshifts of $0.1 < z < 4.5$ on the 2.5 m SDSS telescope with the BOSS spectrograph \citep{Shen2016, Grier2017}. The wavelength coverage of BOSS spectra is $\sim$ 3650–10400\AA\,  with a spectral resolution of $R \sim 2000$. Additional photometric data were acquired with the 3.6 m Canada–France–Hawaii Telescope (CFHT) and the Steward Observatory 2.3 m Bok telescope.

A code called PrepSpec \footnote{PrepSpec: https://github.com/Alymantara/pyPrepSpec} is used to improve the relative flux calibrations and produce line/continuum light curves \citep{Shen2016, Grier2017, Khatu2023}. What PrepSpec does is similar to the decomposition approach used in other studies \citep[e.g., ][]{Bian2010b, Barth2015, Hu2015, Hu2021}. Because we use the light curves derived from PrepSpec, the code is introduced as follows.
For each SDSS-RM spectrum for an RM quasar, the PrepSpec model is: $\mu(\lambda, t)=p(t)[A(\lambda, t)+B(\lambda, t)+C(\lambda, t)]$, where $p(t)$ are time-dependent photometric corrections, $A(\lambda)$ is the average spectrum, and $B (\lambda, t)$ and $C (\lambda, t)$ are variations in the BLR spectrum and continuum, respectively.  $A(\lambda)=\bar {F}(\lambda)+N(\lambda)$, where $\bar{F}(\lambda)$ includes both the continuum and broad-line components, as well as the \feii pseudo-continuum (host-galaxy light is not included ). The narrow-line component, $N(\lambda)$, is isolated using a piecewise cubic spline fit to $A(\lambda)$ as a high-pass filter and then multiplying the result by a window function that is 0 outside and 1 inside a defined velocity range around a specified list of narrow emission lines.  
The BLR variations $B (\lambda, t)$ are represented by a separable function for each line, $B(\lambda, t)=\Sigma_{l=1}^{N_l}B_l(\lambda)L_l(t)$, where the light curve $L_l(t)$ is normalized to a mean of 0 and rms of 1, and $B_l(\lambda)$ is the rms spectrum of line $l$. A key feature of PrepSpec is the inclusion of a time-dependent flux correction calculated by assuming that there is no intrinsic variability of the narrow emission-line fluxes. PrepSpec minimizes the apparent variability of the narrow lines by fitting a model to the spectra that includes intrinsic variations in both the continuum and different broad emission lines.  The narrow-line and broad-line window widths are determined through iterating until convergence in PreSpec. It  is different to other decomposition methods using multi-Gaussians for one by one spectrum \citep[e.g., ][]{Bian2010b, Barth2015, Hu2015, Hu2021}.   

The light curves of \hb, \ha, and the continuum derived from PrepSpec model for 44 quasars ($z<0.53$) are given in the website of Keith Horne \footnote{http://star-www.st-and.ac.uk/$\sim$kdh1/pub/sdss/2020/sdssz.html}. Because the long coverage of light curves (MJD from 2456660 to 2459029, about 6.5 years ) and the updated spectral analysis, we use the data provided by the website of Keith Horne  instead of the data shown in Table 2 in \cite{Grier2017} \footnote{https://cdsarc.cds.unistra.fr/ftp/J/ApJ/851/21/} .  The observation cadence ( average sampling interval of the objects) is large , about 5 days in the years of 2014, about 13 days in years of 2015-2017, and about 20 days in years of 2018-2020.
The sample of these 44 SDSS-RM quasars is used to investigate the Balmer decrement \bd.

\begin{figure}
\centering
\includegraphics[angle=0,width=4in]{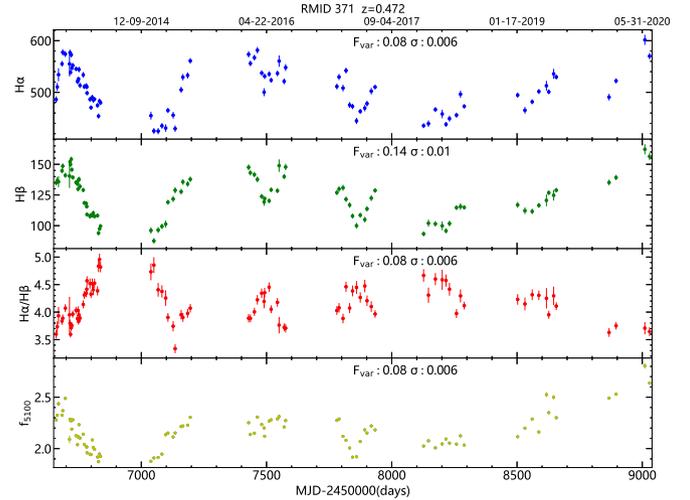}   
\caption{Light curves of the \ha, \hb, \bd, continuum at 5100 \AA\ (from top to bottom) for an example of quasar, SDSS J141123.42+521331.7 (RMID 0371, z = 0.472).
The relative variance and its error for each light curve  are  shown in the upper middle corner of each panel.
}
\label{fig1}
\end{figure}

Figure \ref{fig1} shows an example of the light curve of the Balmer decrement, as well as the \ha, \hb and the continuum at 5100 \AA\ for one quasar (RMID 0371, z = 0.472). 
For the light curve of \bd, the ratio of \ha to \hb emission line flux is calculated from the website of Keith Horne. Through the error propagation formula, errors of \ha and \hb flux are used to calculate the error of \bd, 
\begin{equation}
\sigma=\frac{f_{\rm \ha}}{f_{\rm \hb}}\sqrt{(\sigma_{\rm \hb}/f_{\rm \hb})^2+(\sigma_{\rm \ha}/f_{\rm \ha})^2} 
\end{equation}
where $\sigma$ is the error of Balmer decrement, $f_{\rm \ha}$ and $f_{\rm \hb}$ are the emission fluxes of \ha and \hb, $\sigma_{\rm \ha}$ and $\sigma_{\rm \hb}$ are the corresponding errors of \ha and \hb.

\section{RESULTS AND DISCUSSIONS}

\subsection{The relative variance of the Balmer decrement \bd}

\begin{figure}
\centering
\includegraphics[angle=0,width=3.5in]{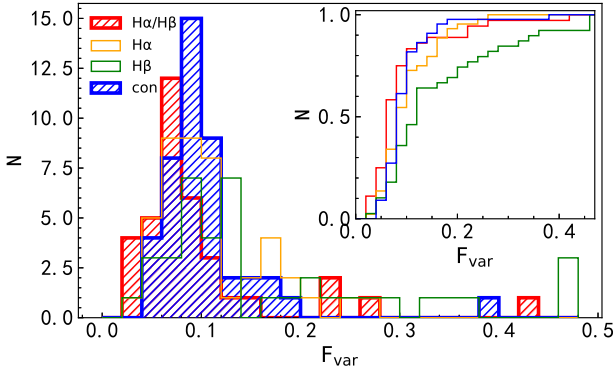}
\caption{Distribution of $F_{\rm var}$ for \bd (filled with red lines), the continuum  at 5100 \AA\ (filled with blue lines), \ha (orange lines) and \hb (green lines). The inserted panel  in the upper right corner shows the corresponding cumulative distribution of $F_{\rm var}$. }
\label{fig2}
\end{figure}

Accounting for the measurement uncertainties, the relative intrinsic variability amplitude $F_{\rm var}$ is used to show the variance in the light curve. $F_{\rm var}$ and its uncertainty $\sigma_{\rm Fvar}$ of a light curve are as commonly defined \citep[e.g., ][]{RP1997, Edelson2002, F2016,  Lu2019a, Zhao2020}:
\begin{equation}
F_{\rm var}=\frac{(\sigma^2-\Delta^2)^{1/2}}{\bar{f}}, 
\sigma_{\rm Fvar}=\frac{1}{F_{\rm var}}(\frac{1}{2N})^{1/2}\frac{\sigma^2}{\bar{f}}, 
\end{equation}
where $\bar{f}$ is the average flux of $f_i$, $\sigma$ is the standard deviation of flux $f_i$, $N$ is the number of observations, $\Delta^2$ is the mean square of uncertainty $\Delta_i$ on the flux $f_i$.  $F_{\rm var}$ gives an estimate of the relative intrinsic variability amplitude by accounting for the measurement uncertainties.

For each light curve,  we calculate the  relative intrinsic variability amplitude $F_{\rm var}$ and its error, where are presented in Table \ref{tab1}. There are 7 quasars without values of $F_{\rm var}$  because of $\Delta^2$ larger than $\sigma^2$.  Figure \ref{fig2} is the $F_{\rm var}$ distribution of the broad-line Balmer decrement \bd, as well as  \ha, \hb and continuum at 5100 \AA\  for our sample.  The mean $F_{\rm var}$ of \bd  is 0.11$\pm$0.019, the mean $F_{\rm var}$ of \ha of all sample is 0.11$\pm$0.009, the mean $F_{\rm var}$ of \hb is 0.22$\pm$0.023,  and the mean $F_{\rm var}$ of continuum is 0.10$\pm$0.008. The mean $F_{\rm var}$ of \bd is similar to \ha and the continuum, but smaller than \hb.

Kolmogorov–Smirnov (K–S) test  \footnote{We used python (scipy.stats.ks\_2samp, scipy.stats.spearmanr) to do our following analysis for the K-S test and the Spearman correlation test.} is performed  to investigate the distribution differences among them. 
For the distributions of the decrement \bd and \ha,  K-S test shows the statistic d and the significant level probability for the null hypothesis ($p$) are 0.23 and 0.20. It shows that there are no obvious difference between their distributions.  It is the case for the distributions of \bd and continuum at 5100 \AA\ with d=0.24 and $p=0.17$ .
For the distributions of \bd and \hb,  d=0.46 and $p=2\times 10^{-4}$, which showing the difference between their distributions.

\subsection{The mean value of \bd and the relation with the SMBH accretion}
\begin{figure}
\centering
\includegraphics[angle=0,width=3.5in]{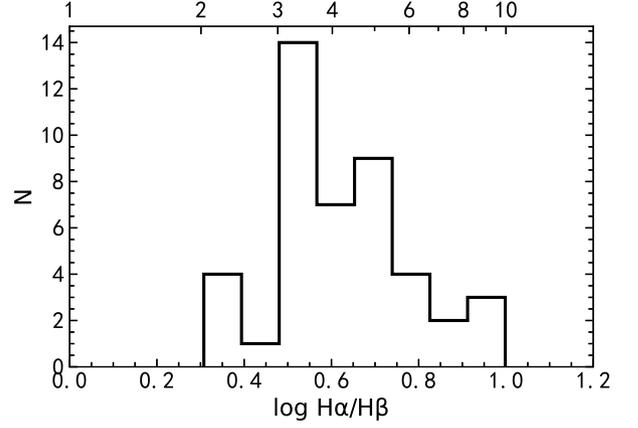}
\caption{The distribution of the mean Balmer decrement ($\log \rm \ha/\hb$) showing a large range from 0.3 to 1.}
\label{fig3}
\end{figure}

\begin{figure*}
\centering
\includegraphics[angle=0,width=3.2in]{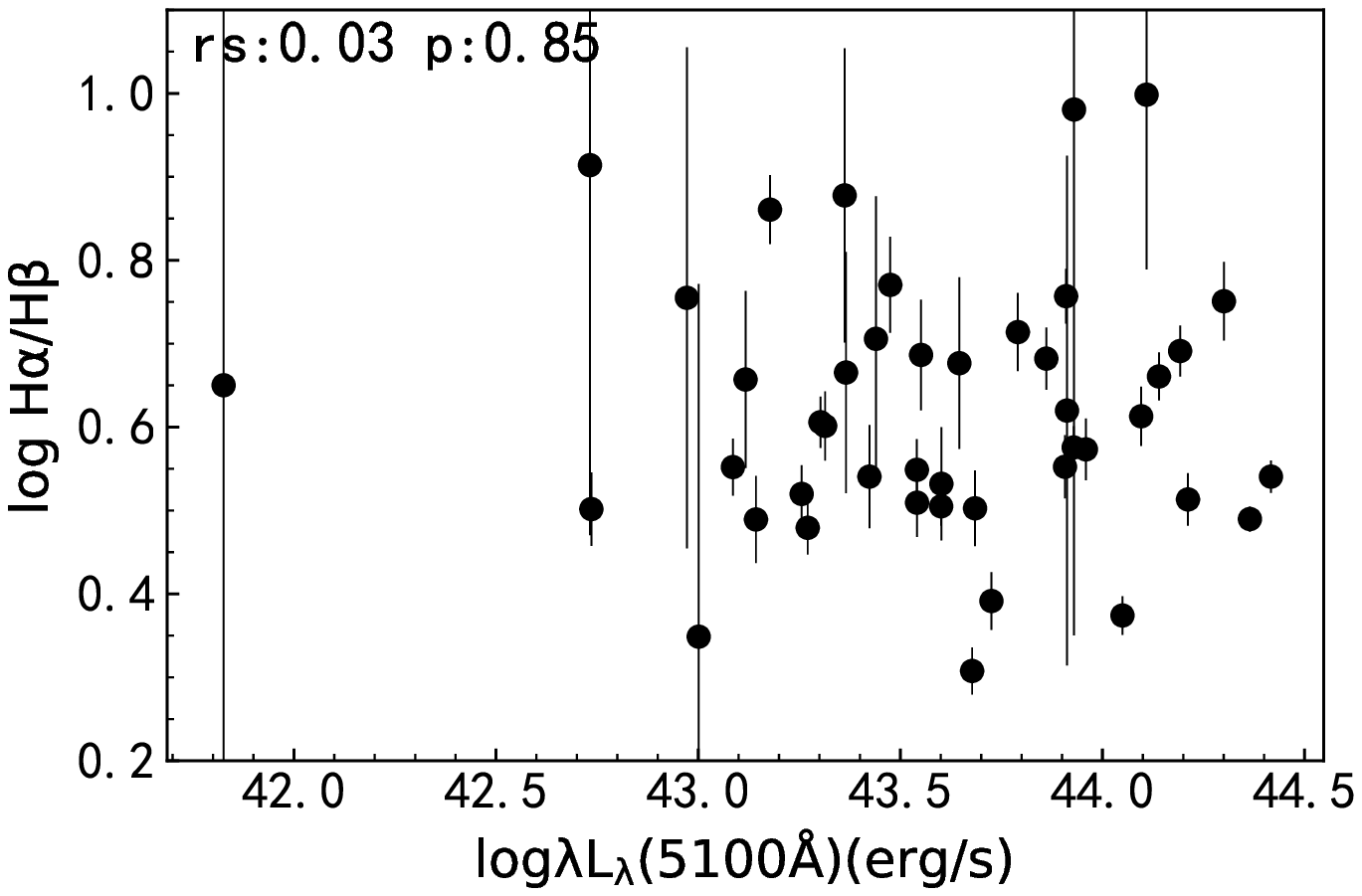}
\includegraphics[angle=0,width=3.2in]{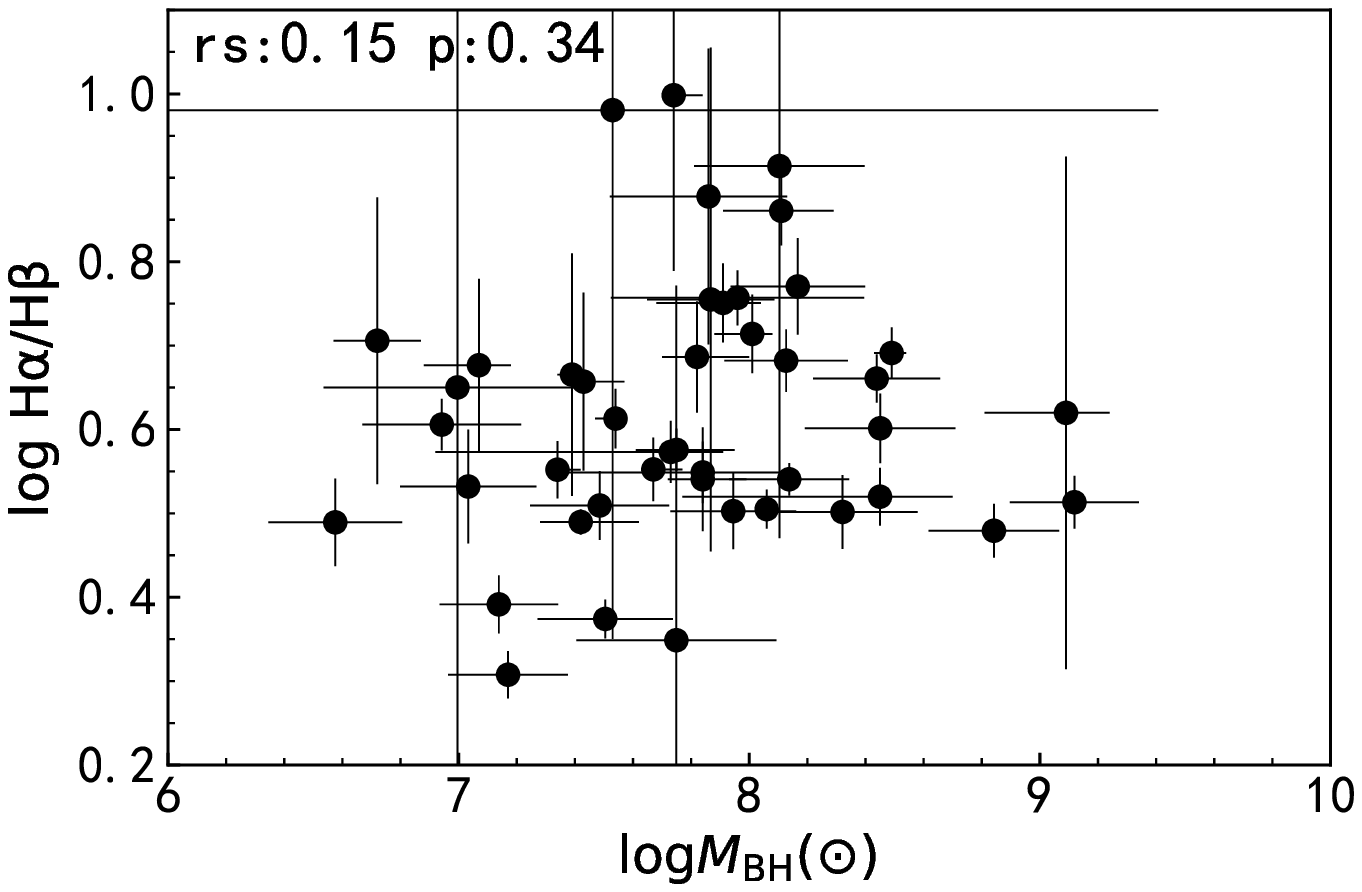}
\includegraphics[angle=0,width=3.2in]{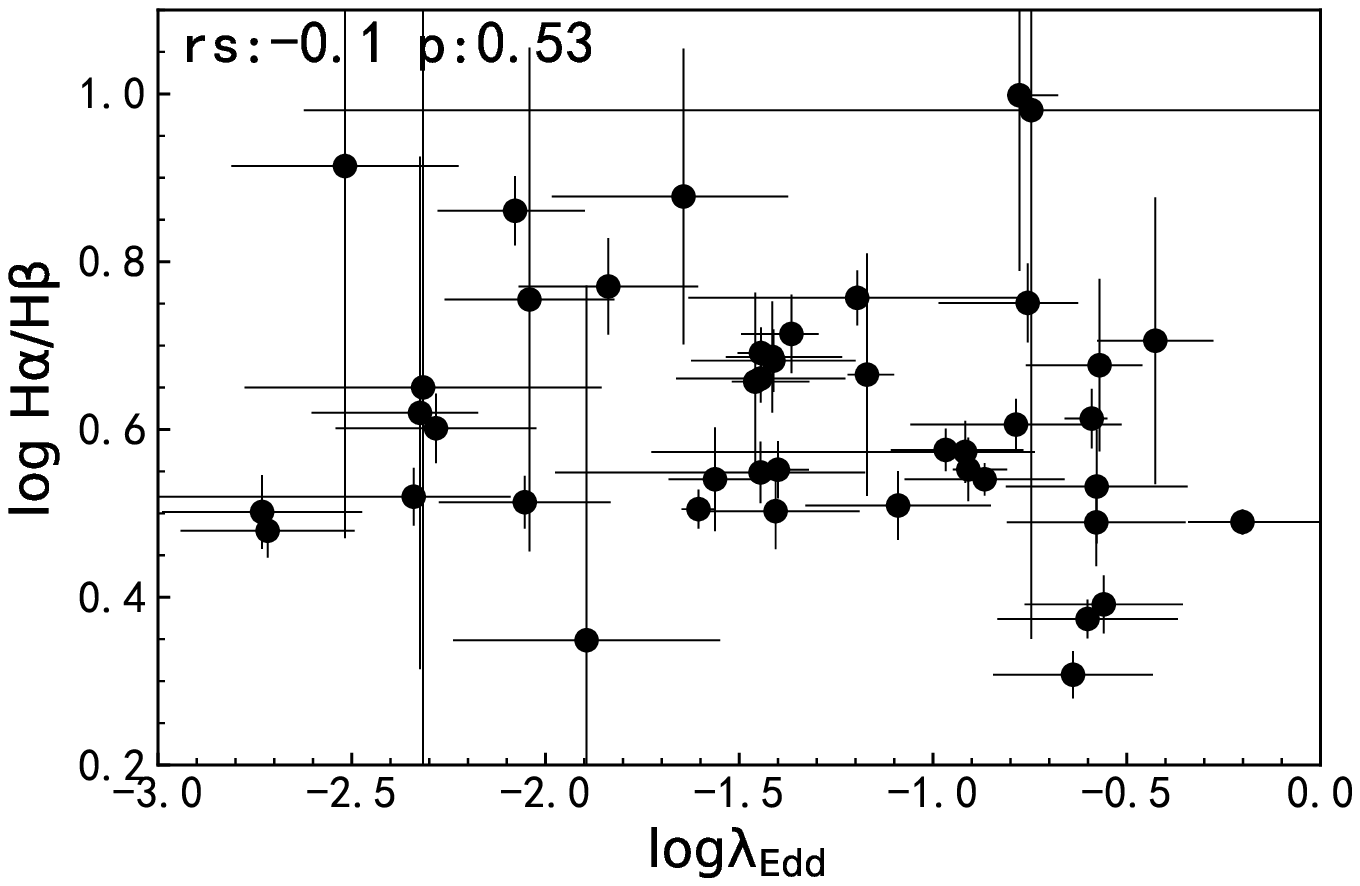}
\includegraphics[angle=0,width=3.2in]{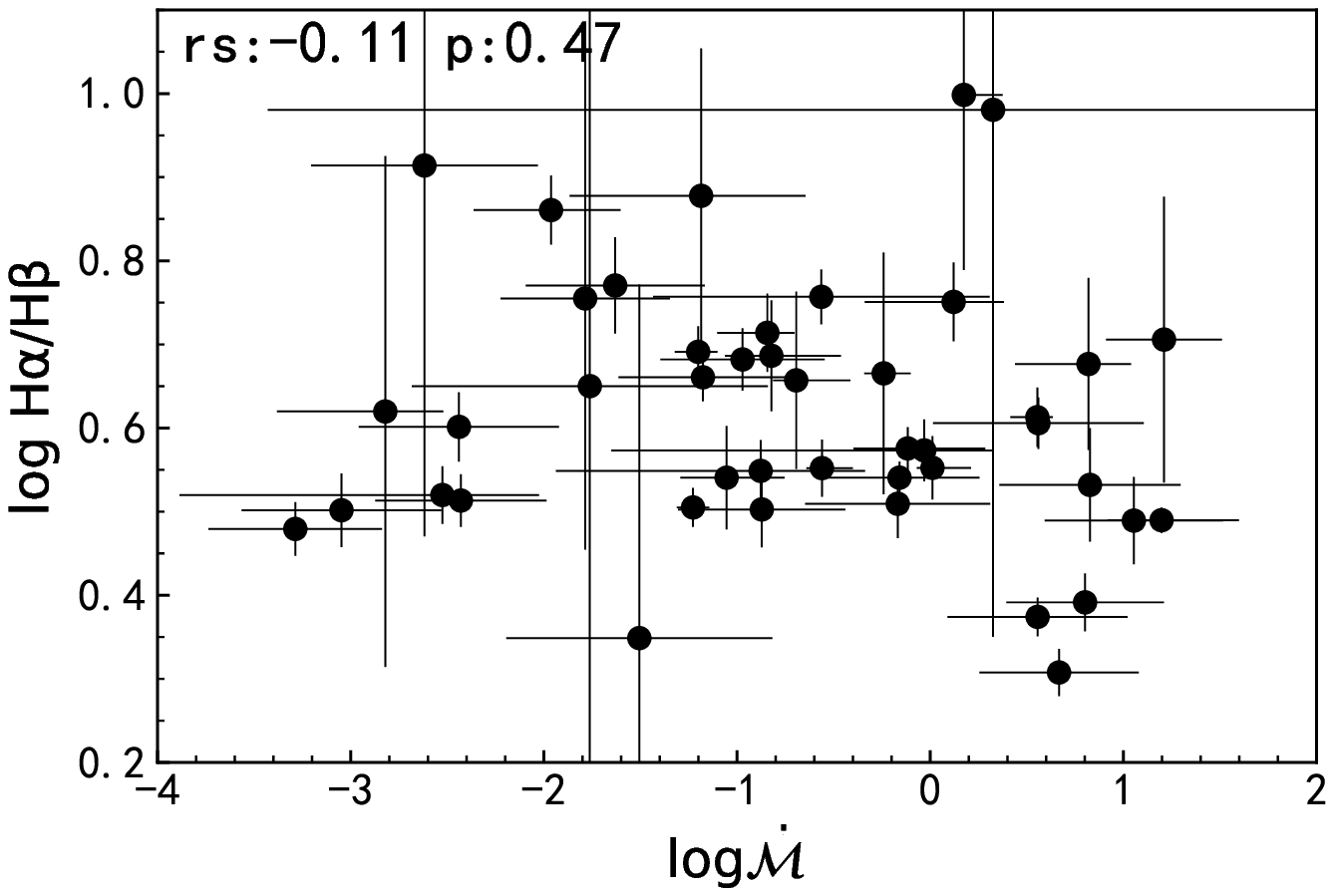}
\caption{The relations of the mean Balmer decrement with the the continuum \lv (top left), the SMBH mass  \mbh (top right), the Eddington ratio \leddR (bottom left) and the dimensionless accretion rate \mdot (bottom right). The Spearman test result is shown in the top corner in each panel. }
\label{fig4}
\end{figure*}

For each quasar, we use the square of the measurement error as the weight to calculate the mean value of \bd. The error of the mean \bd is calculated from the weighted mean of the measurement errors and the weighted standard deviation. 
Figure \ref{fig3} shows the distribution of  the mean Balmer decrement $\log (\bd)$. The mean value of  $\log  (\bd)$ is 0.62 with the standard deviation of 0.15 dex. For $\bd$, the mean value is 4.17 with the standard deviation of 1.42.  We notice that our measured BLR \bd have a larger range extending from 2 to 10.  It is found that the BLR \bd has a strong relation with the equivalent width (EW) of the broad \hb line ($r_s=-0.43, p_{\rm null}=4.0\times 10^{-3}$), while no obvious relation relation with \ha EW ($r_s=0.073, p_{\rm null}=0.64$). It is possible that the large \bd is due to weak \hb line.  Therefore, if we excluding quasars with $\rm \hb(\rm EW) <  20$, the mean  $\log  \bd$ is 0.57 with the standard deviation of 0.11 dex, i.e.,  $3.71\pm 0.90$ for the mean $\bd$.  For BLR \bd, its value changes from one sample to another. \cite{Dong2008} found it is 3.06, \cite{Gaskell2017} suggested it is 2.72 and \cite{Lu2019b} found it is 3.16. However, \cite{Lakicevic2017} found it is  3.76, as well as 3.38 found by \cite{Siram2021}. Our result  is similar to that by \cite{Lakicevic2017}. 

For quasars with reliable lag detection \citep{Grier2017},  \mbh is adopted from col. (11) in Table 1 from \cite{Yu2020b}, as well as its error, which is calculated from the broad \hb line dispersion from the rms spectrum and the lag suggested  by \citep{Grier2017}.
For other quasars, considering the extended empirical \RL relation including \rfe \citep{Du2019, Yu2020a}, \mbh is calculated as follows: $ \log \frac{\mbh}{\msun}=7.01+2\log\frac{\rm FWHM_{\hb}}{1000~\rm km/s}+0.48\log l_{44}-0.38\rfe$ \citep{Liu2022}, where $l_{44}$, $\rm FWHM_{\hb}$ and optical \rfe are measured from the mean spectra \citep{Shen2019}. 
Based on the error transfer formula, the error of \mbh is calculated from the errors of  $\rm FWHM_{\rm \hb}$, \lv, \rfe, including a systematic error of 0.2 dex for $R_{\rm BLR}(\rm \hb)$. The error of \mdot  is about double \mbh error. The error of \leddR is about 0.33 dex \citep[e.g., ][]{Liu2022, Chen2022}. 

In Table \ref{tab1},  \lv, $\rm FWHM_{\rm \hb}$, \rfe, \mbh are shown in cols. (14-17).
\cite{Sulentic2000} proposed two main populations on the plane of $\rm FWHM_{\hb}$ versus \rfe  \citep{Sh14}:  Population A (Pop. A) for quasars with $\rm FWHM_{\hb} < 4000~\kms$ and Population B (Pop. B) for those with  $\rm FWHM_{\hb} > 4000~  \kms$. For our sample, there are 23 quasars belong to Pop. B, 21 quasars belong to Pop. A and  5 quasars belong to extreme Pop. A with $\rfe> 1.2$.

In Figure \ref{fig4}, we show the relations of the mean value of \bd with \lv, \mbh, \leddR and \mdot. There is a large \mbh error bar for RMID 0767, which  is due to its large error of $\rm FWHM_{\hb}$, i.e., $2003\pm 3868~ \kms$.  It seems that there exists a trend of smaller \bd with larger \lv and \leddR or \mdot. Using the Spearman correlation test to investigate the correlations between the mean  Balmer decrement and \lv, \mbh, \leddR, \mdot. The test coefficient $r_s$ and the probability of the null hypothesis $p_{\rm null}$ are 0.03,   0.85 for \lv, 0.15 and 0.34 for \mbh, -0.1 and 0.53 for \leddR, -0.11 and 0.47 for \mdot.  No obvious correlations are found between the mean  Balmer decrement and the \lv, \mbh, \leddR, \mdot.

\subsection{The relation between the Balmer decrement variance and the continuum variance at 5100 \AA}	

\begin{figure}
\centering
\includegraphics[angle=0,width=3.5in]{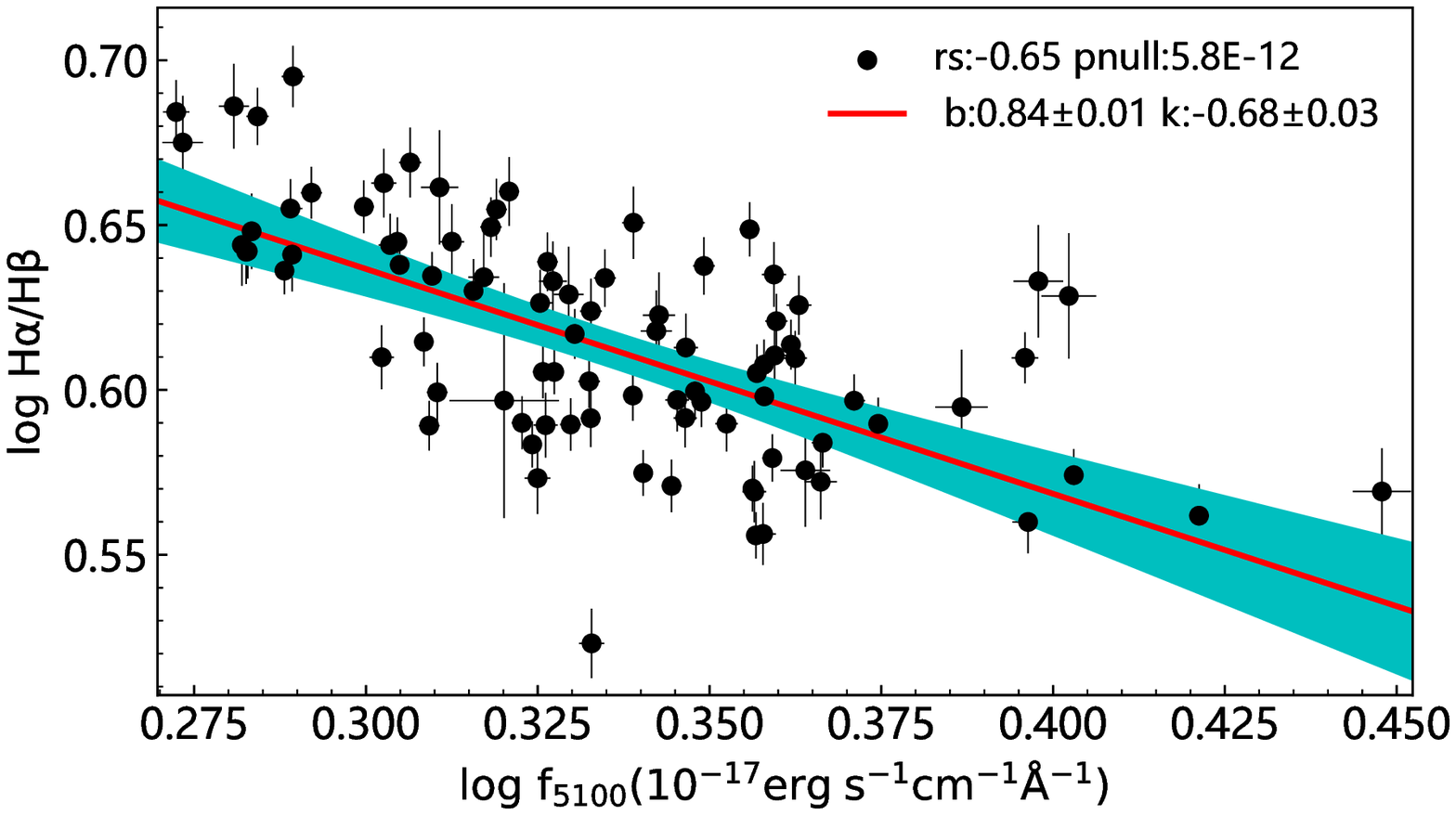}
\includegraphics[angle=0,width=3.5in]{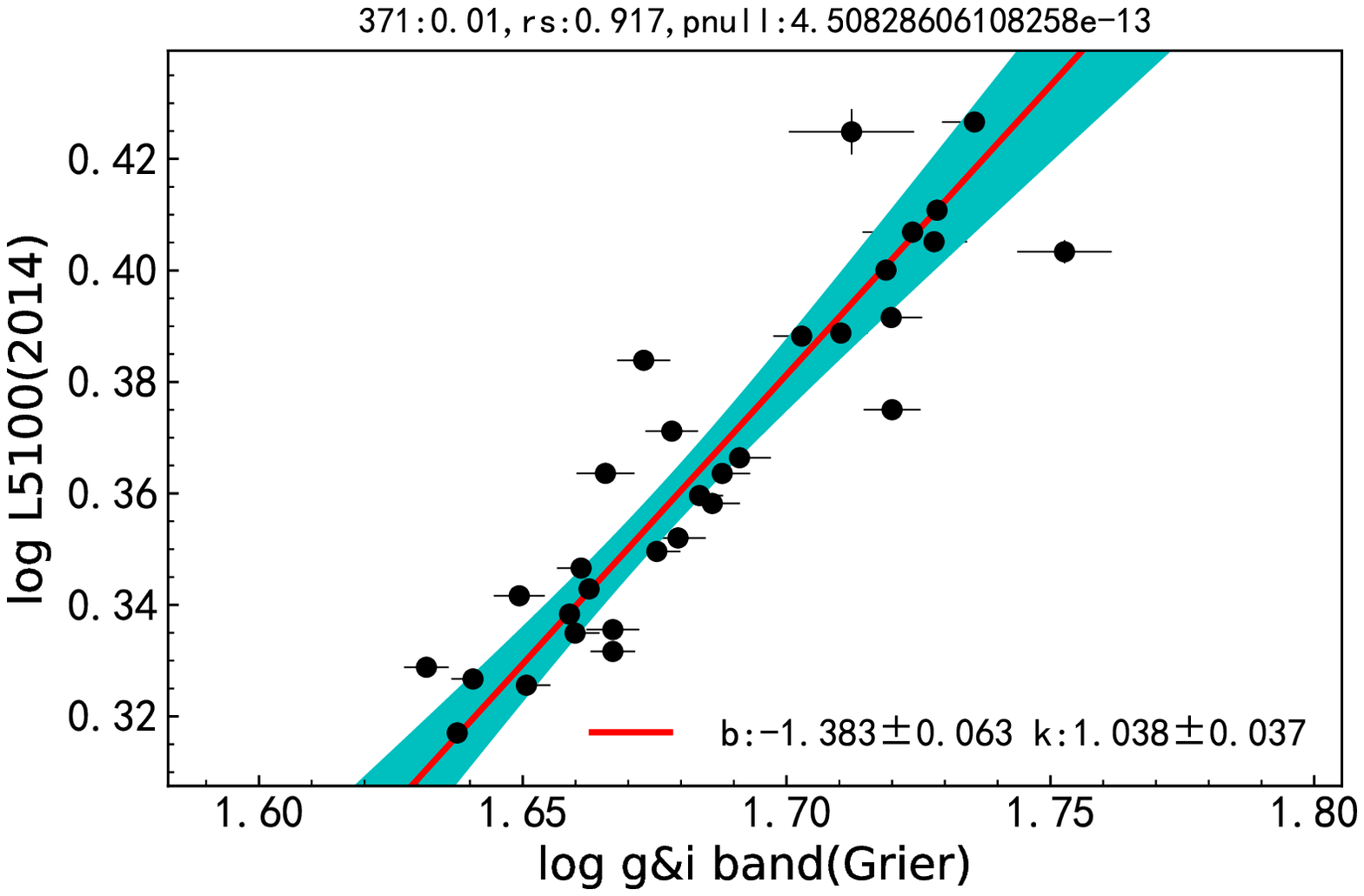} 
\caption{Top: the relationship between  \bd and continuous spectral flux for an example of quasar, SDSS J141123.42+521331.7 (RMID 0371, z = 0.472). The red solid line is the best linear fitting and the blue region is the confidence band.  Bottom:  the continuum at 5100 \AA\ versus the continuum flux at g+i band for the spectra at the year of 2014 \citep{Grier2017} for this quasar. The red line and the blue region are the same as in the  top panel.
}
\label{fig5}
\end{figure}

\begin{figure}
\centering
\includegraphics[angle=0,width=3.5in]{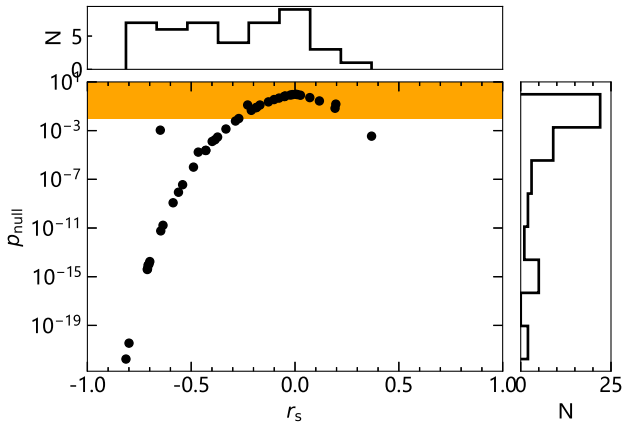}
\caption{The distribution of the  Spearman correlation coefficient $r_{\rm s}$ versus the distribution of probability of the null hypothesis $p_{\rm null}$. The orange shaded region shows $p_{\rm null}$ $\geq$ 0.01,  indicating no strong correlation. }
\label{fig6}
\end{figure}

\begin{figure}
\centering
\includegraphics[angle=0,width=3.9in]{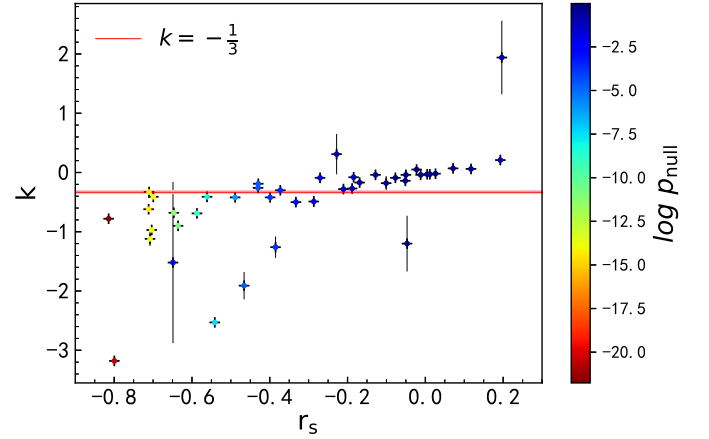}
\caption{The slope versus $r_s$ for the relation between the Balmer decrement and continuum flux at 5100 \AA. The color bar shows the $p_{\rm null}$ from the Spearman test . The red solid line shows the theoretical value of slope ($k=-1/3$) predicted from the extinction and the constant continuum/line flux.}
\label{fig7}
\end{figure}

It was suggested that there existed a correlation between \bd and the continuum for a single AGN \citep[e.g.,][]{Siram2021, RS2022, Li2022}. 	
Figure \ref{fig5} shows an example of the relation between the variance of the Balmer decrement \bd and the variance of the continuum at 5100 \AA\ for one quasar (RMID 0371). There exists a strong correlation with $r_s=-0.65$ and $p_{\rm null}=5.8\times 10^{-12}$.  
Its continuum changes about 0.2 dex, and its decrement changes about 0.18 dex.  Considering the errors in both coordinates, through kmpfit  \footnote {https://www.astro.rug.nl/software/kapteyn/kmpfittutorial.html.},
the best linear fit is $\log \bd = -(0.68 \pm 0.03) log f_{con} +(0.84\pm0.01)$.  The fitting slope is $-0.68\pm 0.03$, which would be test by the dust obscuration (see next paragraph). 
In the bottom panel in Figure \ref{fig5}, we also show the continuum at 5100 \AA\ versus the continuum flux at g+i band for the spectra at the year of 2014 \citep{Grier2017}. The relation between them is strong with $r_s=0.92$. The linear fitting slope by kmpfit is $1.038\pm 0.037$. Although existing line-contribution at g+i band, the slope of 1.038 shows the spectral continuum fluxes  are consistent well with the photometric continuum fluxes at g+i band .

For all 44 quasars in our sample, the correlation coefficient $r_s$, $p_{\rm null}$ and the fitting slope $k$ are given in Cols. (11-13) in Table \ref{tab1}. 
For the $r_s$ distribution, the mean value of the total sample is -0.45 and the standard deviation is 0.78. 
In Figure \ref{fig6}, we show $p_{\rm null}$ versus $r_s$, as well their histograms in the top and right panel. The orange shaded region shows the $p_{\rm null} \ge 0.01$. Considering $p_{\rm null} < 0.01$, there are about 61\% ( 27/44) quasars showing obvious anti-correlation. There is only one quasar with $r_s >0 $ with $p_{\rm null}<0.01$.  For the subsample with  $p_{\rm null} < 0.01$,  the mean  value of  $r_s$ is -0.55  with a standard deviation of  0.16.

It is generally believed that Balmer decrements can be used for dust extinction. The extinction at the wavelength of $\lambda$  can be calculated using the formula: $<A(\lambda)/A(V)>=a(x)+b(x)/R_V$, where $x{\equiv}1/{\lambda} $.  According to the definition of extinction, $A(\lambda)=m^{\rm obs}_{\lambda}-m^{\rm int}_{\lambda}=-2.5 \log\frac{f^{\rm obs}_{\lambda}}{f^{\rm int}_{\lambda}}$, where $m^{\rm obs}_{\lambda}, m^{int}_{\lambda}$ are the observed and intrinsic magnitude, $f^{\rm obs}_{\lambda}, f^{int}_{\lambda}$ are the observed and intrinsic flux.  Therefore,  at the wavelength of  $\lambda=5100~ \AA$:
\begin{equation} 
 f^{int}_{5100}=f^{obs}_{5100} \left [ {\rm \frac{({ \frac{\ha}{\hb}})^{obs}}{({\rm \frac{\ha}{\hb}})^{\rm int}}} \right ]^{ \frac{A(5100 \AA)}{A (\rm \hb)-A(\rm \ha)} }
 \label{eq5} 
 \end{equation}
Using the extinction law of the Milky Way with $R_V=3.1$,  it is found that $A(5100 \AA)=1.11/A_V, A(\rm \ha)=1.18/A_V, A(\rm \hb)=0.81/A_V$.  We obtain the index in Eq. \ref{eq5} is 3. 

Assuming the variance in \bd and the continuum is coming from the dust extinction, i.e., the intrinsic \bd and the continuum
are stable, from Eq. \ref{eq5}, a relation can be found: $\Delta (\log \bd) \propto - \frac{1}{3} \Delta( \log L_{\rm 5100})$, i.e, a slope of $-1/3$. Figure \ref{fig7} shows the slope $k$ versus $r_s$ ,as well as the color bar based on $p_{\rm null}$.  Red solid line shows the line of  $k=-\frac{1}{3}$ predicted from the obscuration  and the pink shaded area shows the 10\% uncertainty.  Considering error of 10\%, there are only 4 quasars consistent with the slope of -1/3.
Most of our measured slopes are not  consistent with the expected value of  $-1/3$ derived from the dust obscuration. It implies that there exists intrinsic variance of the continuum and the line flux. 

\subsection{The lag of the Balmer decrement}
\begin{figure}
\centering
\includegraphics[angle=0,width=3.2in]{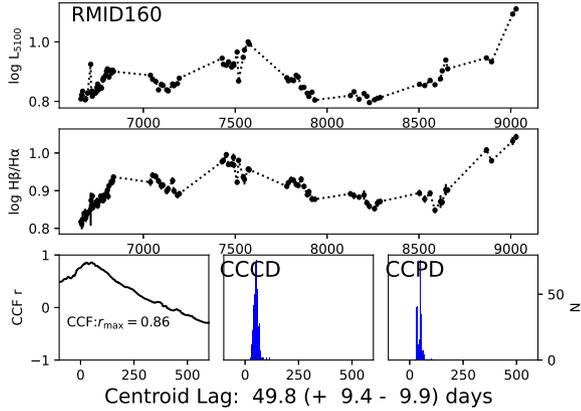}
\caption{Light curves and CCF model for the the inverse of  $\rm \ha/\hb$ analysis for an example of a quasar, SDSS J141041.25+531849.0 (RMID 0160, z = 0.359). The light curves of the continuum ($\log f_{\rm con} $) and the inverse of the broad-line  Balmer decrement ($\log \rm \hb/\ha$) (vertically shifted to assure positive values) are presented in the top and middle panels. For display purposes, we show the weighted mean of all epochs observed within a single night. The bottom three panels show the results of the time-series analysis. The bottom left panel shows the CCF. The other two panels present the lag distributions for the two different methods, normalized to the tallest peak in the distribution. The bottom middle panel shows the CCCD, the bottom right panel shows the CCPD lag distribution. For RMID 0160, the lag is $49.8^{+9.4}_{-9.9}$ light days.} 
\label{fig8}
\end{figure}

\begin{figure}
\centering
\includegraphics[angle=0,width=3.2in]{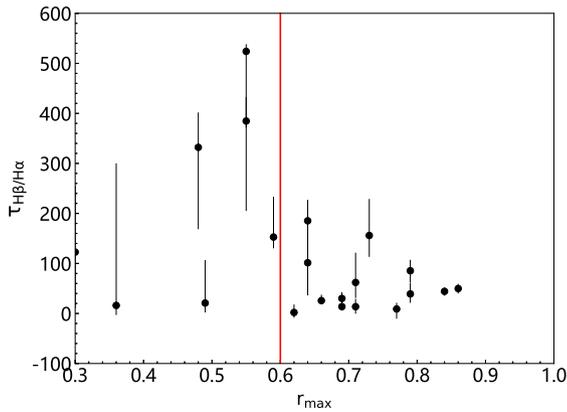}
\caption{ Lag of the Balmer decrement $\tau_{\rm cent} (\rm \hb/\ha)$  versus the maximum correlation coefficient $ r_{\rm max}$ for light curves obtained in years of 2014-2020. The red line shows $r_{\rm max}=0.6$.
}
\label{fig9}
\end{figure}

\begin{figure}
\centering
\includegraphics[angle=0,width=3.2in]{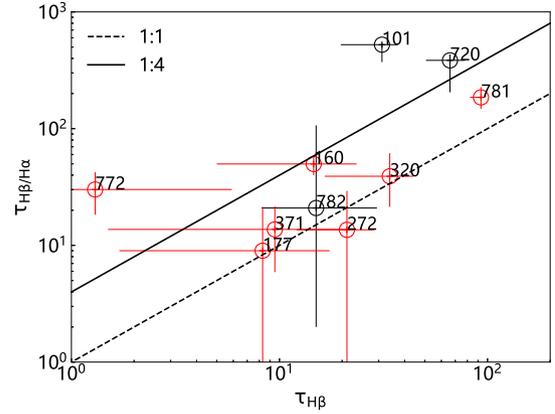}
\includegraphics[angle=0,width=3.2in]{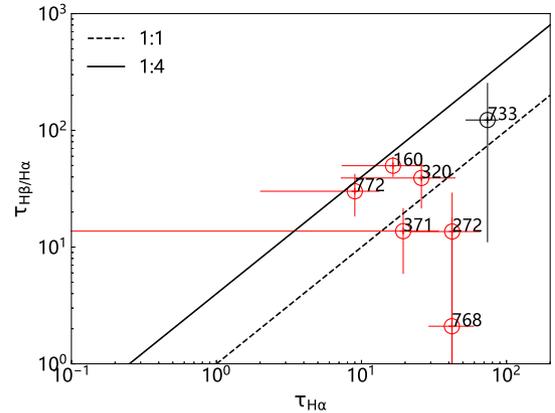}
\caption{Comparison between the \bd lag and the \hb  lag (top), the \ha lag (bottom). 
The dot line is 1:1, and the solid line is 1:4 for the size ratio of BLR to the torus. The red circles denote quasars with $r_{\rm max}> 0.6$. The labels near locations of data points are quasars RMID. }
\label{fig10}
\end{figure}

Because of the anti-correlation between \bd and the continuum, we calculate the light curve of the inverse \bd, i.e.,  $\log \rm \hb/\ha$, and use the common methods to detect the lag with respect to the continuum ($\log f_{\rm con}$). There are mainly three methods to do that,  the interpolated cross-correlation function (ICCF) (e.g., Peterson et al. 2004),  JAVELIN \citep{Zu2011, Zu2013}, and CREAM \citep{Starkey2016}. 
ICCF is the most common method to measure time lags between two light curves in AGN \citep[e.g., ][]{Pe04, Bian2010b, Grier2017}. For 26 quasars with reliable measurement of \hb or \ha lag given by \cite{Grier2017} , from light curves of \bd and the continuum at 5100 \AA,  we calculate CCFs and centroid lag $\tau_{\rm cent}$ using an interpolation grid spacing of 2 days, and a time lag limiting to a range between -100 and 600 days.  The centroid lag is measured using points surrounding the maximum correlation coefficient $r_{\rm max}$ out to $r \ge 0.8 r_{\rm max}$, as is standard for CCF analysis \citep[e.g., ][]{Pe04, Grier2017}.  The lag of the Balmer decrement is listed in col. 22 in Table \ref{tab2}.

Figure \ref{fig8} shows an example about the CCF analysis for the light curves of \hb/\ha and the continuum at 5100 \AA\ for years of 2014-2020  for SDSS J141041.25+531849.0 (RMID 0160, z = 0.359). The light curves of the continuum and \hb/\ha are presented in the top and middle panels. For display purposes, we show the weighted mean of all epochs observed within a single night. The bottom three panels show the results of the time-series analysis. The bottom left panel shows the CCF. With the increasing lag, the correlation coefficient reach the maximum ($ r_{\rm max}=0.86$) at about 49.8 light days, and then slow down with weak correlation. The other two panels present the lag distributions from the two different methods,  cross-correlation centroid distribution ( CCCD) and cross-correlation peak distribution (CCPD).  

Figure \ref{fig9} shows the lag $\tau_{\rm cent}$ of the Balmer decrement versus the maximum correlation coefficient $ r_{\rm max}$ for light curves obtained in years of 2014-2020. The Balmer decrement lags are removed for quasars with $r_{\rm max} < 0.3$ as not significant. 
The maximum ICCF correlation coefficient, $r_{\rm max}$, was suggested to be greater than 0.45. This ensures that the behaviour in the two light curves is well correlated \citep{Grier2017}. Adopting $r_{\rm max} > 0.6$,  there are 14 out 26 quasars showing a reliable measurement of  $\tau_{\rm cent}$.

For SDSS RM quasars, the observation numbers are  about 32, 12, 13, 12, 9, 8, 4  during seven years from 2014 to 2020. Considering the larger gaps between adjacent years (see Figure \ref{fig1}), we also do the ICCF analysis for other two cases, one is for  the spectra observed  during the year of 2014,  and the other is during years of 2014-2015. For our sample, the results on $\tau_{\rm cent}$ and  $r_{\rm max}$ for the Balmer decrement, as well as \hb, \ha from ICCF are given in  cols. (6-11). and cols. (12-17), cols. (18-23) in Table \ref{tab2}  for the  year of 2014, for years of 2014-2015 and for years of 2014-2020, respectively.  
Comparing the decrement lag in three cases, i. e., 2014, 2014-2015, 2014-2020,  it is found that, for the light curves obtained in the year of 2014, the decrement lag has large uncertainty. With all data in the years of 2014 to 2020, the decrement lag has small uncertainty.  It is possible due to small amount of data for ICCF (about 30 data points).  Therefore, we use the decrement lag measurement from the data during seven years from 2014 to 2020 (about 90 data points). 

Balmer decrement is obtained as the ratio of fluxes of \ha and \hb, which have different lags as the response to the continuum light curve.  The curve obtained from the ratio of \ha and \hb  may represent the curve of the one with the larger amplitudes. It is possible that the Balmer decrement lag is false if the decrement lag is  similar to the lag of the \hb or \ha.
Comparing our measured \hb lags with the decrement lags through ICCF from light curves obtained during the years of 2014-2020, it is found that there are ten quasars showing obvious lag difference (considering lag errors).  For the case of our measured \ha lags and the decrement lags through ICCF,  there are 12 quasars showing obvious lag difference  (considering lag errors). The subsample of these 12 quasars is used to investigate the relation between the decrement lag and the BLR lag.

The RM program gave empirical BLR \RL relation \citep{Pe04, Bentz2013, Du2019, Yu2020a}.   \citet{Yang2020} used infrared WISE W1 data to measure the  torus size and gave the torus R-L relation. The torus size is about 4 times of the BLR size. 
Considering our smaller numbers of the continuum data points and large gaps  between adjacent years, we use the lag results for \hb and \ha from \cite{Grier2017}.  
In Figure \ref{fig10},  we compare the Balmer decrement lag and the \hb lag (top), or the \ha lag (bottom).  
Red circles denote quasars with $r_{\rm max} > 0.6$. The solid line is 1:4, i.e., $\tau_{\bd}=4\tau_{\rm \hb}( or~\tau_{\rm \ha})$ and the dash line is 1:1. There are larger uncertainties in the decrement lags, \hb, or \ha lags.
For the top panel in Figure \ref{fig10} , i.e., the decrement lag versus the \hb lag, there are 3 quasars located above the line of 1:4, 6 quasars located between two lines of 1:1 and 1:4, and one quasar located below the line of 1:1.
For the bottom panel in Figure \ref{fig10}, i.e., the decrement lag versus the \ha lag, there is 4 quasars located  between two lines of 1:1 and 1:4, three quasars located below the line of 1:1. 
From our results, the size corresponding to the Balmer decrement lag extends from the BLR size to the torus size.

We use the data from  the spectrum decomposition through PreSpec, the host contribution is not included. It would influence the measurements of the contribution of the AGN continuum, as well as \feii quasi-continuum \citep[e.g., ][]{Bon2020}. This limitations in our data may influence the final results.

\section{Conclusions}
For a sample of 44  SDSS RM quasars ($z<0.53$) with spectral monitoring of  \hb and \ha observed from 2014 to 2020 (~6.5 years), using the light curves of \ha, \hb and the continuum at 5100 \AA\ derived from the spectral decomposition through PrepSpec, the variance of \hb/\ha is investigated. The main conclusions can be summarized as follows:   
\begin{itemize}
\item  The logarithm of the mean broad-line Balmer decrement is 0.62 with a standard deviation of 0.15 dex. 
\item  No obvious correlations are found between the mean Balmer decrement and the SMBH accretion properties, such as the continuum luminosity, the black hole mass,  the Eddington ratio, the dimensionless accretion rate.
\item  There are 27 quasars (61\%) showing strong negative correlations between the Balmer decrement and the continuum, i.e., the Balmer decrement would be smaller with larger continuum flux. Assuming that the dust obscuration leads to the variance in the Balmer decrement and the continuum, the expected slope is -1/3, which is not consistent with most of our measured slopes.
\item  For 26 quasars with reliable measurement of the \hb and/or \ha delay , through ICCF, the time delays between the inverse  Balmer decrement and the continuum are measured. Using the maximum correlation coefficient larger the 0.6, reliable decrement lags are detected for 14 quasars . It suggests that the size of the Balmer decrement lag extends from the BLR size to the torus size.

\end{itemize}

\section*{Acknowledgements}
We are also very grateful to the anonymous referee for her/his instructive comments which significantly improved the content of the paper. This work is supported by the National Key Research and Development Program of China (No. 2017YFA0402703). This work has been supported by the National Science Foundation of China (No. 11973029).

\section*{Data Availability}
The data underlying this article are available at \href{http://star-www.st-and.ac.uk/~kdh1/pub/sdss/2020/sdssz.html}{http://star-www.st-and.ac.uk/~kdh1/pub/sdss/2020/sdssz.html}.

	\clearpage
	
\begin{table*}
\centering
\caption{The relative variability amplitude for \bd, \hb, \ha, \lv, as well as the properties for the relation between \bd and the continuum at 5100 \AA\ and the SMBH masses.
Col. (1): RMID, col. (2): redshift; col. (3): the broad Balmer decrement, col. (4): $F_{\rm var}$ of \ha/\hb, col. (5): the flux of the broad \ha in units of  $\rm 10^{-17}~erg~ s^{-1}~cm^{-2}$,  col. (6):  $F_{\rm var}$ of \ha, col. (7): the flux of the broad \hb in units of $\rm 10^{-17}~erg~s^{-1}~cm^{-2}$, col. (8): $F_{\rm var}$ of \hb, col. (9):  the continuum flux at 5100 \AA\ in $\rm 10^{-17}~erg~ s^{-1} ~cm^{-2} ~\AA^{-1}$, col. (10): $F_{\rm var}$ of the continuum at 5100 \AA, cols. (11-13): $r_s, p_{\rm null}, k$ for the relation between the \bd and the continuum, cols. (14-17) : the luminosity in 5100 \AA\ in units of \ergs, FWHM of broad \hb line in units of \kms, \rfe and $\log \mbh/\msun$. 
}

\renewcommand\arraystretch{1.3}
\tabcolsep=0.05cm
\begin{lrbox}{\tablebox}
\begin{tabular}{ccccccccccccccccc}
\hline
$\rm RMID$  & $z$  & \ha/\hb & $F_{\rm var}(\rm \ha/\hb)$ & \ha & $F_{\rm var}(\rm \ha)$ & \hb & $F_{\rm var}(\rm \hb)$& $f_{5100}$    & $F_{\rm var} (f_{5100})$ & $r_s$ & $p_{\rm null}$ & $k$ &$L_{5100}$ & $\rm  FWHM_{\rm \hb}$ &\rfe  & \mbh \\
(1) & (2) & (3) & (4) & (5) & (6) & (7) & (8) & (9) & (10) & (11) & (12) & (13) & (14) & (15) & (16)& (17) \\
  \hline
        17&0.46&3.58$\pm$0.54&-&343.73$\pm$36.09&0.25$\pm$0.02&54.77$\pm$6.33&0.68$\pm$0.05&5.28$\pm$0.9&0.38$\pm$0.03&-0.81&0.0&-0.78$\pm$0.02&43.91$\pm$0.0&6075.62$\pm$121.48&0.67$\pm$0.01&$9.09^{+0.28}_{-0.15}$\\
        85&0.24&3.0$\pm$0.33&0.07$\pm$0.02&1188.08$\pm$129.16&0.12$\pm$0.01&384.49$\pm$42.82&0.13$\pm$0.01&6.29$\pm$0.76&0.16$\pm$0.01&-0.01&0.91&-0.04$\pm$0.01&43.27$\pm$0.0&13797.44$\pm$363.93&0.26$\pm$0.02&$8.84^{+0.22}_{-0.22}$\\
        88&0.52&4.56$\pm$0.51&0.04$\pm$0.03&497.95$\pm$44.34&0.06$\pm$0.01&107.71$\pm$12.12&0.09$\pm$0.01&2.6$\pm$0.31&0.09$\pm$0.01&-0.37&0.0&-0.3$\pm$0.04&44.14$\pm$0.0&5492.04$\pm$82.07&0.37$\pm$0.03&$8.44^{+0.22}_{-0.22}$\\
        101&0.46&3.08$\pm$0.34&0.03$\pm$0.01&1100.52$\pm$120.82&0.04$\pm$0.0&353.74$\pm$39.09&0.04$\pm$0.0&4.8$\pm$0.58&0.09$\pm$0.01&-0.27&0.01&-0.09$\pm$0.01&44.36$\pm$0.0&2651.7$\pm$302.18&0.43$\pm$0.04&$7.42^{+0.14}_{-0.2}$\\
        126&0.19&4.02$\pm$0.45&0.06$\pm$0.02&461.65$\pm$50.66&0.1$\pm$0.01&113.84$\pm$12.72&0.09$\pm$0.01&7.61$\pm$0.88&0.06$\pm$0.0&-0.05&0.64&-0.04$\pm$0.06&43.3$\pm$0.0&1879.01$\pm$155.38&1.27$\pm$0.03&$6.94^{+0.27}_{-0.27}$\\
        160&0.36&5.09$\pm$0.57&0.1$\pm$0.04&844.94$\pm$89.94&0.16$\pm$0.01&153.99$\pm$17.03&0.25$\pm$0.02&2.15$\pm$0.25&0.15$\pm$0.01&-0.71&0.0&-0.62$\pm$0.02&43.79$\pm$0.0&5034.4$\pm$34.91&0.3$\pm$0.01&$8.01^{+0.13}_{-0.07}$\\
        177&0.48&3.71$\pm$0.44&0.08$\pm$0.03&436.59$\pm$46.5&0.08$\pm$0.01&116.55$\pm$13.31&0.14$\pm$0.01&2.13$\pm$0.27&0.14$\pm$0.01&-0.7&0.0&-0.41$\pm$0.02&43.96$\pm$0.0&5229.7$\pm$76.04&0.47$\pm$0.02&$7.73^{+0.81}_{-0.18}$\\
        184&0.19&2.02$\pm$0.22&0.06$\pm$0.01&830.7$\pm$90.81&0.07$\pm$0.01&405.1$\pm$45.05&0.06$\pm$0.0&12.56$\pm$1.44&0.08$\pm$0.01&-0.13&0.23&-0.04$\pm$0.02&43.68$\pm$0.0&1837.28$\pm$11.06&0.9$\pm$0.01&$7.17^{+0.21}_{-0.21}$\\
        191&0.44&4.58$\pm$0.58&0.23$\pm$0.12&133.34$\pm$15.04&0.22$\pm$0.02&27.61$\pm$3.29&0.33$\pm$0.03&1.59$\pm$0.19&0.1$\pm$0.01&-0.21&0.05&-0.28$\pm$0.08&43.65$\pm$0.01&2177.65$\pm$43.98&0.65$\pm$0.04&$7.07^{+0.19}_{-0.11}$\\
        229&0.47&4.72$\pm$0.56&0.13$\pm$0.06&198.06$\pm$21.75&0.11$\pm$0.01&38.47$\pm$4.3&0.24$\pm$0.02&1.21$\pm$0.14&0.1$\pm$0.01&-0.59&0.0&-0.69$\pm$0.04&43.55$\pm$0.0&4328.84$\pm$297.53&0.6$\pm$0.03&$7.82^{+0.12}_{-0.18}$\\
        252&0.28&3.14$\pm$0.36&0.08$\pm$0.03&352.08$\pm$39.12&0.06$\pm$0.0&109.28$\pm$12.36&0.09$\pm$0.01&2.56$\pm$0.29&0.05$\pm$0.0&-0.29&0.01&-0.49$\pm$0.09&42.74$\pm$0.01&10565.13$\pm$692.7&0.44$\pm$0.03&$8.32^{+0.26}_{-0.26}$\\
        270&0.42&5.84$\pm$0.72&-&217.7$\pm$25.92&0.09$\pm$0.01&37.04$\pm$4.46&0.13$\pm$0.01&1.52$\pm$0.18&0.12$\pm$0.01&-0.17&0.13&-0.17$\pm$0.06&43.48$\pm$0.0&6470.68$\pm$181.01&0.89$\pm$0.05&$8.17^{+0.23}_{-0.23}$\\
        272&0.26&3.75$\pm$0.41&0.06$\pm$0.02&2368.22$\pm$258.11&0.07$\pm$0.0&620.05$\pm$68.34&0.09$\pm$0.01&7.74$\pm$0.92&0.12$\pm$0.01&-0.71&0.0&-0.33$\pm$0.01&43.93$\pm$0.02&3514.91$\pm$17.43&0.33$\pm$0.01&$7.75^{+0.14}_{-0.2}$\\
        305&0.53&4.89$\pm$0.55&0.06$\pm$0.03&504.81$\pm$56.88&0.07$\pm$0.01&102.54$\pm$11.64&0.08$\pm$0.01&2.59$\pm$0.32&0.08$\pm$0.01&0.01&0.91&-0.03$\pm$0.03&44.19$\pm$0.0&2917.6$\pm$61.7&0.47$\pm$0.04&$8.49^{+0.06}_{-0.05}$\\
        320&0.26&3.41$\pm$0.43&0.14$\pm$0.04&509.16$\pm$57.2&0.11$\pm$0.01&145.64$\pm$16.99&0.22$\pm$0.02&3.69$\pm$0.43&0.1$\pm$0.01&-0.7&0.0&-0.97$\pm$0.04&43.42$\pm$0.0&4700.09$\pm$54.96&0.67$\pm$0.02&$7.84^{+0.12}_{-0.15}$\\
        338&0.42&6.83$\pm$1.06&0.08$\pm$2.53&138.23$\pm$16.67&0.11$\pm$0.01&15.45$\pm$1.93&0.37$\pm$0.04&1.59$\pm$0.19&0.07$\pm$0.01&-0.47&0.0&-1.91$\pm$0.23&43.36$\pm$0.0&6789.45$\pm$579.67&0.0$\pm$0.0&$7.86^{+0.34}_{-0.27}$\\
        341&0.42&3.47$\pm$0.4&0.04$\pm$0.01&1199.25$\pm$137.62&0.08$\pm$0.01&347.85$\pm$39.71&0.1$\pm$0.01&5.53$\pm$0.69&0.1$\pm$0.01&-0.43&0.0&-0.19$\pm$0.02&44.42$\pm$0.0&3471.49$\pm$24.98&0.57$\pm$0.01&$8.14^{+0.21}_{-0.21}$\\
        371&0.47&4.07$\pm$0.47&0.08$\pm$0.03&506.45$\pm$56.22&0.08$\pm$0.01&122.14$\pm$13.75&0.14$\pm$0.01&2.13$\pm$0.24&0.08$\pm$0.01&-0.65&0.0&-0.68$\pm$0.03&44.1$\pm$0.0&4105.97$\pm$37.6&0.75$\pm$0.02&$7.54^{+0.07}_{-0.04}$\\
        377&0.34&4.49$\pm$0.63&0.22$\pm$0.46&148.29$\pm$16.57&0.17$\pm$0.01&30.46$\pm$3.71&0.34$\pm$0.03&2.44$\pm$0.28&0.05$\pm$0.0&-0.39&0.0&-1.26$\pm$0.18&43.37$\pm$0.0&4429.12$\pm$105.45&0.95$\pm$0.05&$7.39^{+0.05}_{-0.07}$\\
        497&0.51&3.24$\pm$0.37&0.06$\pm$0.02&612.98$\pm$69.21&0.07$\pm$0.01&186.55$\pm$21.43&0.03$\pm$0.0&3.88$\pm$0.46&0.08$\pm$0.01&0.0&0.96&-0.03$\pm$0.03&44.21$\pm$0.0&11162.28$\pm$264.83&0.21$\pm$0.02&$9.12^{+0.22}_{-0.22}$\\
        518&0.46&5.68$\pm$0.64&0.04$\pm$0.04&266.66$\pm$29.53&0.05$\pm$0.0&46.56$\pm$5.24&0.07$\pm$0.01&1.24$\pm$0.14&0.09$\pm$0.01&-0.18&0.08&-0.08$\pm$0.05&43.91$\pm$0.0&3321.98$\pm$900.62&0.0$\pm$0.0&$7.96^{+0.44}_{-0.44}$\\
        541&0.44&3.21$\pm$0.37&0.03$\pm$0.05&122.05$\pm$13.74&0.11$\pm$0.01&37.49$\pm$4.28&0.11$\pm$0.01&1.08$\pm$0.13&0.12$\pm$0.01&0.19&0.07&0.21$\pm$0.08&43.54$\pm$0.0&3245.29$\pm$111.17&1.52$\pm$0.07&$7.49^{+0.24}_{-0.24}$\\
        645&0.47&8.94$\pm$1.44&0.52$\pm$0.76&453.81$\pm$50.61&0.06$\pm$0.0&37.09$\pm$4.6&0.47$\pm$0.04&2.28$\pm$0.26&0.1$\pm$0.01&-0.8&0.0&-3.18$\pm$0.08&44.11$\pm$0.0&4259.19$\pm$90.47&0.4$\pm$0.01&$7.74^{+0.03}_{-0.1}$\\
        720&0.47&5.58$\pm$0.66&0.11$\pm$0.05&677.02$\pm$77.4&0.08$\pm$0.01&119.07$\pm$13.81&0.14$\pm$0.01&3.68$\pm$0.45&0.08$\pm$0.01&-0.4&0.0&-0.42$\pm$0.03&44.3$\pm$0.0&3405.89$\pm$22.33&0.52$\pm$0.01&$7.91^{+0.23}_{-0.13}$\\
        733&0.46&4.77$\pm$0.55&0.07$\pm$0.04&389.41$\pm$44.72&0.1$\pm$0.01&80.97$\pm$9.26&0.1$\pm$0.01&1.96$\pm$0.23&0.11$\pm$0.01&-0.02&0.83&0.05$\pm$0.03&43.86$\pm$0.0&4689.79$\pm$44.47&0.6$\pm$0.02&$8.13^{+0.21}_{-0.21}$\\
        766&0.16&2.44$\pm$0.28&0.08$\pm$0.02&1494.08$\pm$165.3&0.07$\pm$0.01&601.35$\pm$67.42&0.09$\pm$0.01&21.19$\pm$2.75&0.07$\pm$0.01&0.12&0.27&0.06$\pm$0.02&43.73$\pm$0.0&1936.26$\pm$7.28&1.45$\pm$0.01&$7.14^{+0.2}_{-0.2}$\\
        767&0.53&8.81$\pm$2.73&-&110.39$\pm$25.11&0.02$\pm$0.0&8.26$\pm$2.13&0.5$\pm$0.12&1.02$\pm$0.23&0.07$\pm$0.01&-0.65&0.0&-1.52$\pm$1.36&43.93$\pm$0.0&2003.55$\pm$3868.39&0.0$\pm$0.0&$7.53^{+1.88}_{-1.88}$\\
        768&0.26&3.95$\pm$0.47&0.09$\pm$0.03&959.57$\pm$105.6&0.17$\pm$0.01&233.07$\pm$26.53&0.18$\pm$0.01&6.21$\pm$0.81&0.1$\pm$0.01&-0.49&0.0&-0.42$\pm$0.02&43.31$\pm$0.0&9093.06$\pm$611.77&0.53$\pm$0.02&$8.45^{+0.26}_{-0.26}$\\
        769&0.19&5.25$\pm$0.85&-&249.68$\pm$28.19&0.16$\pm$0.01&37.75$\pm$4.82&0.5$\pm$0.04&5.51$\pm$0.72&0.09$\pm$0.01&-0.71&0.0&-1.12$\pm$0.12&42.97$\pm$0.0&6250.34$\pm$124.45&1.05$\pm$0.02&$7.87^{+0.22}_{-0.22}$\\
        772&0.25&4.52$\pm$0.66&0.42$\pm$0.23&454.49$\pm$49.76&0.12$\pm$0.01&74.52$\pm$8.66&0.46$\pm$0.03&5.48$\pm$0.63&0.07$\pm$0.01&-0.54&0.0&-2.53$\pm$0.06&43.44$\pm$0.0&3089.67$\pm$65.88&0.67$\pm$0.02&$6.72^{+0.15}_{-0.15}$\\
        775&0.17&3.5$\pm$0.39&0.08$\pm$0.02&1489.45$\pm$166.07&0.08$\pm$0.01&411.82$\pm$45.64&0.13$\pm$0.01&13.54$\pm$1.66&0.06$\pm$0.0&-0.19&0.08&-0.27$\pm$0.03&43.54$\pm$0.0&3662.02$\pm$26.73&0.73$\pm$0.01&$7.84^{+0.53}_{-0.27}$\\
        776&0.12&4.39$\pm$0.56&0.28$\pm$0.11&1275.48$\pm$140.8&0.11$\pm$0.01&265.29$\pm$30.33&0.29$\pm$0.02&12.09$\pm$1.48&0.12$\pm$0.01&-0.64&0.0&-0.9$\pm$0.03&43.12$\pm$0.0&4344.84$\pm$557.6&0.78$\pm$0.04&$7.43^{+0.06}_{-0.14}$\\
        779&0.15&3.53$\pm$0.41&0.07$\pm$0.02&767.31$\pm$86.9&0.1$\pm$0.01&213.23$\pm$24.35&0.09$\pm$0.01&4.81$\pm$0.57&0.08$\pm$0.01&-0.05&0.64&-0.14$\pm$0.02&43.09$\pm$0.0&2944.79$\pm$20.28&0.57$\pm$0.01&$7.34^{+0.04}_{-0.08}$\\
        781&0.26&3.19$\pm$0.36&0.05$\pm$0.01&570.41$\pm$63.34&0.06$\pm$0.0&178.32$\pm$20.09&0.08$\pm$0.01&4.59$\pm$0.56&0.09$\pm$0.01&-0.43&0.0&-0.26$\pm$0.02&43.6$\pm$0.02&2989.96$\pm$48.32&0.59$\pm$0.04&$8.06^{+0.04}_{-0.04}$\\
        782&0.36&3.54$\pm$0.41&0.03$\pm$0.07&444.06$\pm$52.26&0.18$\pm$0.01&124.07$\pm$14.35&0.21$\pm$0.02&4.4$\pm$0.63&0.17$\pm$0.01&-0.56&0.0&-0.41$\pm$0.02&43.91$\pm$0.0&3428.09$\pm$36.56&1.03$\pm$0.02&$7.67^{+0.04}_{-0.1}$\\
        789&0.42&3.14$\pm$0.36&0.09$\pm$0.03&263.38$\pm$30.0&0.07$\pm$0.01&82.37$\pm$9.38&0.12$\pm$0.01&1.55$\pm$0.18&0.08$\pm$0.01&-0.33&0.0&-0.5$\pm$0.05&43.68$\pm$0.0&4148.68$\pm$70.17&0.55$\pm$0.02&$7.94^{+0.22}_{-0.22}$\\
        790&0.24&3.29$\pm$0.38&0.08$\pm$0.02&770.02$\pm$88.43&0.11$\pm$0.01&235.63$\pm$26.97&0.12$\pm$0.01&6.57$\pm$0.82&0.07$\pm$0.01&-0.08&0.47&-0.09$\pm$0.03&43.26$\pm$0.0&9207.55$\pm$268.85&0.66$\pm$0.04&$8.45^{+0.68}_{-0.25}$\\
        792&0.53&6.54$\pm$0.27&-&50.53$\pm$9.13&0.09$\pm$0.02&3.82$\pm$4.2&0.47$\pm$0.09&0.87$\pm$0.25&0.09$\pm$0.01&-0.05&0.7&-1.2$\pm$0.47&43.0$\pm$0.01&5111.22$\pm$823.91&0.81$\pm$0.09&$7.75^{+0.34}_{-0.34}$\\
        797&0.24&2.21$\pm$0.36&0.09$\pm$0.02&79.38$\pm$47.06&0.08$\pm$0.01&35.47$\pm$15.43&0.14$\pm$0.01&2.11$\pm$0.48&0.05$\pm$0.0&-0.1&0.36&-0.18$\pm$0.12&43.14$\pm$0.0&1474.1$\pm$47.1&1.7$\pm$0.03&$6.58^{+0.23}_{-0.23}$\\
        798&0.42&3.08$\pm$0.26&0.04$\pm$0.01&407.92$\pm$18.47&0.06$\pm$0.01&131.24$\pm$8.19&0.05$\pm$0.0&3.69$\pm$0.53&0.08$\pm$0.01&0.07&0.51&0.07$\pm$0.03&44.05$\pm$0.0&2311.19$\pm$62.02&1.13$\pm$0.06&$7.5^{+0.23}_{-0.23}$\\
        822&0.29&2.32$\pm$0.38&0.1$\pm$0.03&185.84$\pm$50.08&0.27$\pm$0.02&74.47$\pm$17.02&0.26$\pm$0.02&4.46$\pm$0.66&0.16$\pm$0.01&0.03&0.81&-0.02$\pm$0.07&43.6$\pm$0.0&1847.81$\pm$63.39&1.47$\pm$0.05&$7.03^{+0.23}_{-0.23}$\\
        840&0.24&3.37$\pm$1.48&0.09$\pm$0.03&534.45$\pm$11.65&0.19$\pm$0.01&152.26$\pm$1.24&0.17$\pm$0.01&5.21$\pm$0.41&0.12$\pm$0.01&0.37&0.0&0.18$\pm$0.02&43.18$\pm$0.0&7409.47$\pm$113.24&1.05$\pm$0.02&$8.11^{+0.2}_{-0.18}$\\
        845&0.27&5.57$\pm$1.5&-&83.07$\pm$6.54&0.17$\pm$0.02&7.06$\pm$0.57&0.72$\pm$0.11&2.63$\pm$0.11&0.1$\pm$0.01&0.2&0.15&1.94$\pm$0.62&42.73$\pm$0.01&9023.24$\pm$930.44&0.87$\pm$0.07&$8.1^{+0.29}_{-0.29}$\\
        846&0.23&3.48$\pm$0.93&-&13.72$\pm$2.17&0.18$\pm$0.03&1.85$\pm$0.34&0.49$\pm$0.12&0.34$\pm$0.05&0.19$\pm$0.02&-0.23&0.13&0.31$\pm$0.34&41.83$\pm$0.01&3464.89$\pm$1038.81&0.0$\pm$0.0&$7.0^{+0.46}_{-0.46}$\\
        \hline
    \end{tabular}%
    \label{tab1}%
\end{lrbox}
\scalebox{0.8}{\usebox{\tablebox}}
\end{table*}%

\clearpage
\newpage

\begin{landscape}
\begin{table}
\centering
\caption{The lags of the broad-line Balmer decrement \hb/\ha,  \hb, \ha for light curves obtained during the year of 2014, years of 2014-2015, years of 2014-2020, as well as the \hb, \ha lags  during the year of 2014  given by \protect \cite{Grier2017}. 
Col. (1): RMID, cols. (2-5): \hb lag, quality, \ha lag, quality, cols. (6-11): lag and $r_{\rm max}$ obtained from the data during the year of 2014 for \hb, \ha, \hb/\ha, respectively, cols. (12-17): lag and $r_{\rm max}$ obtained from the data during the years of 2014-2015 for \hb, \ha, \hb/\ha, respectively, col. (18-23): lag and $r_{\rm max}$ obtained from the data during the years of 2014-2020 for \hb, \ha, \hb/\ha, respectively. The lag is in units of light days in the observed frame. The lines in bold denote quasars shown in Figure \ref{fig10}.
}
\renewcommand\arraystretch{1.3}
\tabcolsep=0.1cm
\begin{lrbox}{\tablebox}
 \begin{tabular}{|c|ccc|cccccc|cccccc|cccccccc|}
 \hline
$\rm RMID$ &$\tau^a(\rm \hb)$ &$q^a(\rm \hb)$& $\tau^a(\rm \ha)$&$q^a(\rm \ha)$&$\tau^b(\rm \hb)$&$r^b(\rm \hb)$&$\tau^b(\rm \ha)$&$r^b(\rm \ha)$&$\tau^b(\rm \hb/\ha)$&$r^b(\rm \hb/\ha)$&$\tau^c(\rm \hb)$&$r^c(\rm  \hb)$&$\tau^c(\rm \ha)$&$r^c (\rm \ha)$&$\tau^c(\rm \hb/\ha)$&$r^c(\rm \hb/\ha)$&$\tau^d(\rm \hb)$&$r^d(\rm \hb)$&$\tau^d(\rm \ha)$&$r^d(\rm \ha)$&$\tau^d(\rm \hb/\ha)$&$r^d(\rm \hb/\ha)$&\\
\hline
&\multicolumn{4}{c}{Grier} &\multicolumn{6}{c}{2014}&\multicolumn{6}{c}{2014-2015}&\multicolumn{6}{c}{2014-2020} \\
(1) & (2) & (3) & (4) & (5) & (6) & (7) & (8) & (9) & (10) & (11) & (12) & (13) & (14) & (15) & (16)& (17) & (18) & (19) & (20) & (21) & (22) & (23) \\
 \hline
        17&$32.7^{+15.4}_{-15.9}$&4.0&$65.3^{+20.0}_{-15.4}$&5.0&$36.0^{+22.9}_{-34.1}$&0.61&$53.9^{+24.4}_{-24.7}$&0.69&$24.3^{+30.7}_{-23.4}$&0.62&$47.9^{+21.8}_{-16.3}$&0.81&$54.0^{+18.7}_{-23.5}$&0.8&$51.0^{+15.9}_{-21.6}$&0.78&$85.8^{+28.3}_{-25.5}$&0.81&$111.3^{+29.0}_{-25.0}$&0.8&$85.4^{+21.9}_{-23.1}$&0.79&\\
          88&-&-&$84.0^{+5.7}_{-8.3}$&3.0&$76.1^{+6.9}_{-158.2}$&0.8&$0.0^{+86.0}_{-82.9}$&0.81&$26.9^{+20.1}_{-85.9}$&0.67&$19.1^{+68.7}_{-20.0}$&0.65&$0.9^{+119.9}_{-1.9}$&0.8&$44.1^{+226.9}_{-17.2}$&0.5&$56.9^{+54.6}_{-54.7}$&0.79&$70.2^{+96.0}_{-70.1}$&0.69&$62.0^{+59.5}_{-31.0}$&0.71&\\
       \bf 101&  $\bf 31.1^{+5.9}_{-11.4}$&\bf 5.0&-&-&$\bf 85.9^{+5.2}_{-65.0}$&\bf 0.72& $\bf 91.0^{+3.1}_{-168.0}$&\bf 0.6& $\bf 65.0^{+7.1}_{-46.0}$&\bf 0.64& $\bf 71.3^{+11.6}_{-19.4}$&\bf 0.9&$\bf 165.1^{+112.9}_{-35.7}$&\bf 0.7&$\bf 26.8^{+44.3}_{-20.3}$&\bf 0.66& $\bf 152.9^{+27.3}_{-18.3}$&\bf 0.8& $\bf 194.4^{+36.1}_{-37.7}$&\bf 0.71&$\bf 523.9^{+14.4}_{-152.2}$&\bf 0.55&\\
\bf        160&$\bf 14.6^{+8.9}_{-9.6}$&\bf 3.0&$\bf 16.5^{+10.3}_{-9.2}$&\bf 4.0&$\bf -68.0^{+77.3}_{-7.5}$&\bf 0.94& $\bf 8.7^{+17.2}_{-75.1}$&\bf 0.82&$\bf -68.1^{+85.0}_{-8.8}$&\bf 0.93&$\bf 71.2^{+127.6}_{-17.4}$&\bf 0.75&$\bf 146.3^{+7.9}_{-37.4}$&\bf 0.83& $\bf 49.1^{+144.3}_{-20.0}$&\bf 0.82& $\bf 100.3^{+11.5}_{-9.7}$&\bf 0.81&$\bf 130.3^{+23.0}_{-20.0}$&\bf 0.65&\bf  $ \bf 49.8^{+9.4}_{-9.9}$&\bf 0.86&   \\
\bf        177&$\bf 8.3^{+9.1}_{-6.6}$&\bf 4.0&-&-& $\bf -0.9^{+6.0}_{-87.1}$&\bf 0.71& $\bf 2.1^{+42.9}_{-83.1}$\bf &\bf 0.65& $\bf -14.9^{+36.3}_{-64.1}$&\bf 0.58&$\bf 16.6^{+16.3}_{-10.0}$&\bf 0.94&$\bf 22.7^{+17.5}_{-9.4}$&\bf 0.88& $\bf 12.2^{+22.9}_{-21.2}$&\bf 0.86&$\bf 28.4^{+11.6}_{-8.5}$&\bf 0.9& $\bf 40.8^{+15.0}_{-11.0}$&\bf 0.79&  $\bf \bf 9.0^{+12.4}_{-19.4}$&\bf 0.77&\\
        191&$14.0^{+5.7}_{-5.8}$&5.0&$26.0^{+9.6}_{-11.8}$&4.0&$34.9^{+3.2}_{-80.0}$&0.64&$37.4^{+19.7}_{-39.3}$&0.43&$34.0^{+16.7}_{-84.1}$&0.58&$117.1^{+10.1}_{-11.7}$&0.71&$41.3^{+97.5}_{-40.3}$&0.55&$120.2^{+7.9}_{-12.4}$&0.64&$106.0^{+27.2}_{-29.4}$&0.77&$110.5^{+28.3}_{-29.2}$&0.76&$101.4^{+56.7}_{-65.2}$&0.64&\\
        229&$21.0^{+6.3}_{-8.7}$&5.0&$34.0^{+19.1}_{-11.5}$&3.0&$28.1^{+14.0}_{-34.2}$&0.57&$44.8^{+10.2}_{-67.9}$&0.54&$15.0^{+22.9}_{-83.0}$&0.52&$111.1^{+19.1}_{-95.0}$&0.46&$134.0^{+111.8}_{-94.9}$&0.55&$112.9^{+15.2}_{-98.7}$&0.51&$23.2^{+26.3}_{-7.5}$&0.68&$28.7^{+126.8}_{-25.7}$&0.67&$25.7^{+11.8}_{-5.2}$&0.66&\\
        252&-&-&$14.1^{+8.1}_{-6.7}$&5.0&$69.0^{+21.9}_{-142.7}$&0.54&$24.0^{+11.9}_{-64.2}$&0.55&$70.0^{+17.1}_{-144.1}$&0.55&$86.1^{+7.0}_{-145.4}$&0.51&$-6.4^{+102.4}_{-37.5}$&0.34&$86.0^{+59.0}_{-157.2}$&0.42&$15.0^{+288.1}_{-5.2}$&0.42&$460.6^{+46.3}_{-246.0}$&0.26&$16.1^{+284.0}_{-19.2}$&0.36&\\
\bf        272&$\bf 21.1^{+7.5}_{-9.0}$&\bf 5.0&$\bf 42.2^{+24.0}_{-21.1}$&\bf 3.0&$\bf 26.3^{+6.0}_{-4.1}$&\bf 0.82&$\bf 50.0^{+13.9}_{-36.0}$&\bf 0.59&$\bf 21.0^{+12.7}_{-11.0}$&\bf 0.79&$\bf 27.4^{+4.6}_{-5.6}$&\bf 0.9&$\bf 47.7^{+21.3}_{-21.8}$&\bf 0.59&$\bf 17.8^{+7.3}_{-6.8}$&\bf 0.83&$\bf 32.2^{+9.4}_{-9.9}$&\bf 0.84&$49.0^{+11.2}_{-10.1}$&\bf 0.65&$\bf 13.6^{+15.7}_{-14.1}$&\bf 0.71&\\
        305&$-74.0^{+22.2}_{-12.8}$&2.0&-&-&$38.8^{+17.8}_{-133.4}$&0.68&$-67.0^{+101.2}_{-8.1}$&0.48&$41.0^{+29.0}_{-132.0}$&0.57&$131.7^{+105.1}_{-75.9}$&0.36&$0.0^{+138.0}_{-75.0}$&0.44&$155.0^{+119.9}_{-92.2}$&0.39&$179.1^{+23.0}_{-27.9}$&0.73&$38.5^{+179.6}_{-30.0}$&0.61&$152.8^{+80.6}_{-22.5}$&0.59&\\
\bf         320&$\bf 33.9^{+10.1}_{-17.4}$&\bf 4.0&$\bf 25.9^{+18.7}_{-18.7}$&\bf 4.0&$\bf 37.0^{+2.9}_{-32.1}$&\bf 0.72&$\bf 17.1^{+21.8}_{-16.1}$&\bf 0.56&$\bf 39.2^{+37.8}_{-94.5}$&\bf 0.69&$\bf 39.0^{+118.1}_{-27.0}$&\bf 0.7&$\bf 8.8^{+40.3}_{-18.7}$&\bf 0.52&$\bf 40.1^{+113.0}_{-39.9}$&\bf 0.56&$\bf 66.2^{+32.7}_{-36.9}$&\bf 0.86&$\bf 97.4^{+24.9}_{-38.5}$&\bf 0.78& $\bf 39.1^{+22.5}_{-17.7}$&\bf 0.79&\\
        338&$10.9^{+14.2}_{-12.8}$&4.0&-&-&$24.9^{+69.0}_{-92.0}$&0.73&$88.0^{+10.0}_{-163.5}$&0.83&$19.6^{+72.4}_{-86.7}$&0.71&$-8.9^{+35.0}_{-47.1}$&0.54&$23.1^{+77.8}_{-55.1}$&0.45&$-8.8^{+35.7}_{-47.2}$&0.53&$-5.1^{+13.7}_{-47.8}$&0.51&$15.4^{+8.9}_{-12.9}$&0.52&$-12.2^{+23.1}_{-44.8}$&0.51&\\
\bf        371&$\bf 9.5^{+12.9}_{-8.0}$&\bf 3.0&$\bf 19.4^{+14.8}_{-20.8}$&\bf 3.0&$\bf 28.5^{+8.0}_{-9.3}$&\bf 0.89&$\bf 29.8^{+10.7}_{-9.6}$&\bf 0.89&$\bf 29.9^{+18.1}_{-18.3}$&\bf 0.9&$\bf 20.1^{+6.1}_{-4.0}$&\bf 0.92&$\bf 41.4^{+9.9}_{-6.2}$&\bf 0.86&$\bf -2.8^{+11.3}_{-11.9}$&\bf 0.82&$\bf 25.1^{+4.0}_{-3.2}$&\bf 0.83&$\bf 38.7^{+8.0}_{-6.4}$&\bf 0.77& $\bf 13.7^{+7.9}_{-7.8}$&\bf 0.69&\\
        377&$12.0^{+16.0}_{-15.5}$&3.0&$7.1^{+14.3}_{-44.0}$&3.0&$21.1^{+69.0}_{-60.0}$&0.5&$16.9^{+11.3}_{-54.9}$&0.66&$24.9^{+68.1}_{-105.9}$&0.51&$13.9^{+10.5}_{-30.3}$&0.57&$19.1^{+15.9}_{-21.9}$&0.68&$9.1^{+115.9}_{-50.1}$&0.46&$339.1^{+60.1}_{-317.0}$&0.51&$350.9^{+53.4}_{-337.2}$&0.51&$332.1^{+69.8}_{-163.5}$&0.48&\\
        645&$7.5^{+9.5}_{-12.5}$&4.0&$26.7^{+9.4}_{-22.9}$&5.0&$28.5^{+12.8}_{-16.2}$&0.93&$29.0^{+15.6}_{-20.3}$&0.86&$32.4^{+10.6}_{-18.1}$&0.93&$41.7^{+10.5}_{-9.4}$&0.91&$31.7^{+7.9}_{-8.1}$&0.88&$43.6^{+10.3}_{-10.2}$&0.91&$44.3^{+7.1}_{-7.6}$&0.85&$39.3^{+7.2}_{-6.2}$&0.83&$44.9^{+7.2}_{-8.7}$&0.84&\\
\bf        720&$\bf 66.0^{+11.9}_{-15.4}$&\bf 2.0&-&-&$\bf 0.0^{+4.9}_{-2.0}$&\bf 0.79&$\bf 24.4^{+68.6}_{-80.6}$&\bf 0.51&$\bf 2.9^{+55.0}_{-10.4}$&\bf 0.64&$\bf 0.0^{+3.8}_{-2.8}$&\bf 0.81&$\bf 0.0^{+10.0}_{-2.1}$&\bf 0.65&$\bf 4.2^{+56.8}_{-10.0}$&\bf 0.56&$\bf 3.0^{+381.0}_{-4.0}$&\bf 0.75&$\bf 1.0^{+371.7}_{-2.9}$&\bf 0.7&$384.7^{+47.9}_{-179.8}$&\bf 0.55&\\
 \bf       733&-&-&$\bf 74.0^{+13.9}_{-21.8}$&\bf 2.0&$\bf -0.9^{+46.0}_{-53.0}$&\bf 0.4&$\bf 44.0^{+14.9}_{-97.0}$&\bf 0.42&$\bf 10.1^{+79.9}_{-28.3}$&\bf 0.57&$\bf 119.8^{+122.2}_{-73.7}$&\bf 0.61&$\bf 102.2^{+47.1}_{-51.1}$&\bf 0.55&$\bf 103.0^{+182.0}_{-127.7}$&\bf 0.34&$\bf 44.1^{+103.8}_{-37.1}$&\bf 0.53&$\bf 11.4^{+109.8}_{-11.4}$&\bf 0.48&$\bf 122.7^{+132.5}_{-111.7}$&\bf 0.3&\\
\bf        768&-&-&$\bf 42.0^{+17.8}_{-13.0}$&\bf 5.0&$\bf 28.0^{+5.0}_{-5.0}$&\bf 0.59&$\bf 36.5^{+23.5}_{-5.4}$&\bf 0.35&$\bf 20.9^{+5.1}_{-27.8}$&\bf 0.63&$\bf 27.4^{+6.0}_{-5.6}$&\bf 0.83&$\bf 30.8^{+26.1}_{-9.7}$&\bf 0.69&$\bf 27.4^{+77.4}_{-15.1}$&\bf 0.8&$\bf 36.2^{+13.0}_{-9.1}$&\bf 0.88&$\bf 67.8^{+31.0}_{-14.7}$&\bf 0.76&$\bf 2.1^{+15.9}_{-10.5}$&\bf 0.62&\\
\bf        772&$\bf 1.3^{+4.6}_{-5.2}$&\bf 5.0&$\bf 9.0^{+4.3}_{-7.0}$&\bf 5.0&$\bf -59.0^{+70.9}_{-3.9}$&\bf 0.52&$\bf 14.0^{+25.0}_{-59.9}$&\bf 0.35&$\bf -60.0^{+69.9}_{-3.1}$&\bf 0.6&$\bf 190.1^{+97.9}_{-175.5}$&\bf 0.53&$\bf 189.9^{+96.0}_{-151.6}$&\bf 0.5&$\bf 189.9^{+97.9}_{-178.9}$&\bf 0.53&$\bf 38.8^{+9.5}_{-11.1}$&\bf 0.69&$\bf 60.6^{+19.0}_{-10.2}$&\bf 0.7&$\bf 30.1^{+12.3}_{-11.8}$&\bf 0.69&\\
        775&$22.1^{+10.2}_{-12.9}$&4.0&-&-&$31.1^{+4.8}_{-4.9}$&0.86&$48.0^{+7.1}_{-12.9}$&0.75&$23.2^{+6.8}_{-15.3}$&0.73&$28.1^{+6.9}_{-4.8}$&0.82&$43.0^{+149.9}_{-13.4}$&0.74&$24.9^{+167.0}_{-5.7}$&0.74&$26.9^{+9.6}_{-8.0}$&0.17&$35.1^{+12.2}_{-35.5}$&0.0&$25.0^{+51.6}_{-9.1}$&0.29&\\
        776&$10.0^{+6.3}_{-4.2}$&4.0&$8.0^{+5.4}_{-6.3}$&4.0&$2.9^{+4.3}_{-20.1}$&0.6&$-26.0^{+25.1}_{-30.0}$&0.6&$7.4^{+4.6}_{-9.2}$&0.6&$154.1^{+39.4}_{-94.1}$&0.51&$68.2^{+158.6}_{-11.3}$&0.54&$150.0^{+59.0}_{-28.1}$&0.49&$137.7^{+67.9}_{-22.9}$&0.8&$130.7^{+21.6}_{-19.9}$&0.85&$155.9^{+73.0}_{-42.7}$&0.73&\\
       779&$12.1^{+9.5}_{-10.1}$&4.0&$16.9^{+21.6}_{-13.7}$&2.0&$19.7^{+7.4}_{-7.5}$&0.89&$14.3^{+25.5}_{-15.3}$&0.71&$66.0^{+19.0}_{-51.5}$&0.75&$29.6^{+11.0}_{-5.7}$&0.78&$41.9^{+141.1}_{-32.0}$&0.52&$55.7^{+13.3}_{-27.4}$&0.65&$23.0^{+7.3}_{-4.6}$&0.69&$21.7^{+6.8}_{-8.8}$&0.6&$326.0^{+136.1}_{-38.7}$&0.28&\\
 \bf       781&$\bf 93.0^{+3.1}_{-10.5}$&\bf 2.0&-&-&$\bf 1.0^{+85.1}_{-80.0}$&\bf 0.51&$\bf 64.1^{+10.9}_{-147.1}$&\bf 0.6&$\bf 0.0^{+19.0}_{-75.2}$&\bf 0.46&$\bf 38.3^{+23.7}_{-28.6}$&\bf 0.89&$\bf 46.7^{+25.5}_{-35.6}$&\bf 0.77&$\bf 22.1^{+63.8}_{-18.9}$&\bf 0.76&$\bf 116.9^{+40.8}_{-39.4}$&\bf 0.72&$\bf 38.2^{+46.6}_{-18.7}$&\bf 0.58& $\bf 185.3^{+41.8}_{-37.3}$&\bf 0.64&\\
 \bf       782&$\bf 15.0^{+14.4}_{-6.8}$&\bf 4.0&-&-&$\bf -73.1^{+165.2}_{-11.9}$&\bf 0.55&$\bf -76.0^{+163.1}_{-10.0}$&\bf 0.32&$\bf 88.9^{+5.0}_{-161.6}$&\bf 0.63&$\bf 3.9^{+112.5}_{-7.9}$&\bf 0.97&$\bf 2.0^{+140.4}_{-7.0}$&\bf 0.97&$\bf 104.8^{+174.8}_{-96.8}$&\bf 0.43&$\bf 5.1^{+80.6}_{-8.4}$&\bf 0.96&$\bf 3.8^{+128.1}_{-9.6}$&\bf 0.96&$\bf 20.9^{+85.8}_{-18.9}$&\bf 0.49&\\
        790&$11.0^{+6.1}_{-6.5}$&3.0&$0.9^{+11.9}_{-5.6}$&2.0&$-1.0^{+7.2}_{-19.3}$&0.58&$-0.9^{+41.0}_{-19.0}$&0.59&$0.1^{+74.8}_{-32.1}$&0.42&$0.0^{+246.2}_{-15.9}$&0.54&$1.9^{+265.3}_{-2.9}$&0.49&$26.0^{+78.2}_{-50.1}$&0.52&$127.2^{+77.1}_{-104.5}$&0.75&$182.6^{+40.7}_{-25.1}$&0.78&$28.9^{+56.3}_{-47.2}$&0.11&\\
        840&$8.1^{+3.4}_{-2.3}$&5.0&$11.9^{+5.6}_{-3.7}$&5.0&$36.2^{+10.8}_{-27.3}$&0.7&$2.9^{+21.0}_{-92.0}$&0.76&$86.1^{+4.8}_{-101.1}$&0.62&$39.6^{+11.5}_{-14.5}$&0.89&$32.0^{+12.7}_{-15.5}$&0.91&$164.1^{+85.8}_{-28.1}$&0.1&$17.1^{+16.6}_{-6.1}$&0.84&$25.9^{+16.3}_{-10.9}$&0.88&$353.9^{+146.9}_{-187.6}$&0.1&\\
        \hline
    \end{tabular}%
    \label{tab2}%
\end{lrbox}
\scalebox{0.70}{\usebox{\tablebox}}
\end{table}%
\end{landscape}

\end{document}